\documentclass[journal]{IEEEtran}
\usepackage{footnote} % footnotes!
\usepackage{hyperref}
\hypersetup{%
    pdfborder = {0 0 0}
}
\usepackage{dblfloatfix}

\newcommand{\red}[1]{\noindent\textit{\color{red}\textbf{}}}
\newcommand{\blue}[1]{\color{black}~#1} % REBUTTAL MODIFICATIONS BLACK

\usepackage[]{siunitx}
\DeclareSIUnit{\degree}{°}
\DeclareSIUnit{\deg}{deg}
\DeclareSIUnit{\nothing}{\relax}
\DeclareSIUnit{\pixel}{px}
\DeclareSIUnit{\fps}{frame/s}
\DeclareSIUnit{\mac}{MAC}

\usepackage{color}
\usepackage{tabularx}
\usepackage{tabularray}
\usepackage{array}
\usepackage{tablefootnote}
\usepackage{booktabs}
\usepackage[flushleft]{threeparttable}
\UseTblrLibrary{diagbox}
\usepackage{cite}

\ifCLASSINFOpdf
  \usepackage[pdftex]{graphicx}
\else
  \usepackage[dvips]{graphicx}
\fi

\usepackage{amsmath,bm} % math symbols, and bold math symbols
\usepackage{utfsym} % checkmarks

\begin{document}

\title{Distilling Tiny and Ultra-fast Deep Neural Networks for Autonomous Navigation on Nano-UAVs}

\author{
% % Authors % <-this % stops a space
Lorenzo Lamberti$^{1}$, 
Lorenzo Bellone$^{2}$, 
Luka Macan$^{1}$,
Enrico Natalizio$^{2,3}$, \\
Francesco Conti$^{1}$, 
Daniele Palossi$^{4,5}$,
and~Luca Benini$^{1,5}$
\thanks{$^{1}$L. Lamberti, L. Macan,  F. Conti, and L. Benini are with the Department of Electrical, Electronic and Information Engineering, University of Bologna, Italy
(email: lorenzo.lamberti@unibo.it, luka.macan@unibo.it, f.conti@unibo.it, luca.benini@unibo.it).
}%
\thanks{$^{2}$L. Bellone and Enrico Natalizio are with the Autonomous Robotics Research Center, Technology Innovation Institute, UAE
(email: lorenzo.bellone@tii.ae, enrico.natalizio@tii.ae).
}%
\thanks{$^{3}$ Enrico Natalizio is also with CNRS, LORIA, University of Lorraine, France
(email: enrico.natalizio@loria.fr).
}%
\thanks{$^{4}$D. Palossi is with the Dalle Molle Institute for Artificial Intelligence, USI-SUPSI, Switzerland.
}%
\thanks{$^{5}$D. Palossi and L. Benini are also with the Integrated Systems Laboratory, ETH Z\"urich, Switzerland
(email: lbenini@iis.ee.ethz.ch).
}
% \thanks{Manuscript received April 19, 2005; revised August 26, 2015.}
}

% The paper headers
% \markboth{IEEE internet of things journal,,~Vol.~x, No.~x, July~2024}%
\markboth{Accepted for publication at IEEE Internet of Things Journal, July~2024}%
{Shell \MakeLowercase{\textit{et al.}}: Bare Demo of IEEEtran.cls for IEEE Journals}

% If you want to put a publisher's ID mark on the page you can do it like
% this:
%\IEEEpubid{0000--0000/00\$00.00~\copyright~2015 IEEE}
% Remember, if you use this you must call \IEEEpubidadjcol in the second
% column for its text to clear the IEEEpubid mark.

\maketitle

\begin{abstract}
Nano-sized unmanned aerial vehicles (UAVs) are ideal candidates for flying Internet-of-Things smart sensors to collect information in narrow spaces.
This requires ultra-fast navigation under very tight memory/computation constraints.
The PULP-Dronet convolutional neural network (CNN) enables autonomous navigation running aboard a nano-UAV at \SI{19}{\fps}, at the cost of a large memory footprint of \SI{320}{\kilo\byte} -- and with drone control in complex scenarios hindered by the disjoint training of collision avoidance and steering capabilities.
In this work, we distill a novel family of CNNs with better capabilities than PULP-Dronet, but memory footprint reduced by up to 168$\times$ (down to \SI{2.9}{\kilo\byte}), achieving an inference rate of up to \SI{139}{\fps}; we collect a new open-source unified collision/steering \SI{66}{\kilo\nothing} images dataset for more robust navigation; and we perform a thorough in-field analysis of both PULP-Dronet and our tiny CNNs running on a commercially available nano-UAV.
Our tiniest CNN, called Tiny-PULP-Dronet v3, navigates with a 100\% success rate a challenging and never-seen-before path, composed of a narrow obstacle-populated corridor and a \SI{180}{\degree} turn, at a maximum target speed of \SI{0.5}{\meter/\second}.
In the same scenario, the SoA PULP-Dronet consistently fails despite having 168$\times$ more parameters.
\end{abstract}


\section*{Supplementary material}
\noindent Supplementary video at: \url{http://youtu.be/ehNlDyhsVSc}.

\noindent Open-source code and dataset: \url{http://github.com/pulp-platform/pulp-dronet}.

\IEEEpeerreviewmaketitle

\section{Introduction} \label{sec:introduction}

\begin{table*}
\centering
\label{tab:rw}
\caption{Related works based on CNNs, running aboard nano-UAVs. The task is either classification (C), or regression (R).
}
\begin{threeparttable}
\begin{tblr}{
  width = \linewidth,
  colspec = {Q[180]Q[100]Q[162]Q[100]Q[90]Q[112]Q[90]Q[90]},
  row{1} = {c},
  cell{2}{2} = {c},
  cell{2}{3} = {c},
  cell{2}{4} = {c},
  cell{2}{5} = {c},
  cell{2}{6} = {c},
  cell{2}{7} = {c},
  cell{2}{8} = {c},
  cell{3}{2} = {c},
  cell{3}{3} = {c},
  cell{3}{4} = {c},
  cell{3}{5} = {c},
  cell{3}{6} = {c},
  cell{3}{7} = {c},
  cell{3}{8} = {c},
  cell{4}{2} = {c},
  cell{4}{3} = {c},
  cell{4}{4} = {c},
  cell{4}{5} = {c},
  cell{4}{6} = {c},
  cell{4}{7} = {c},
  cell{4}{8} = {c},
  cell{5}{2} = {c},
  cell{5}{3} = {c},
  cell{5}{4} = {c},
  cell{5}{5} = {c},
  cell{5}{6} = {c},
  cell{5}{7} = {c},
  cell{5}{8} = {c},
  cell{6}{2} = {c},
  cell{6}{3} = {c},
  cell{6}{4} = {c},
  cell{6}{5} = {c},
  cell{6}{6} = {c},
  cell{6}{7} = {c},
  cell{6}{8} = {c},
  cell{7}{2} = {c},
  cell{7}{3} = {c},
  cell{7}{4} = {c},
  cell{7}{5} = {c},
  cell{7}{6} = {c},
  cell{7}{7} = {c},
  cell{7}{8} = {c},
  hline{1,8} = {-}{0.08em},
  hline{2} = {-}{},
}
\textbf{Work} & \textbf{Device} & \textbf{Description} & \textbf{Task} & \textbf{Power [mW]} & \textbf{Parameters} & \textbf{Operations [MAC]} & \textbf{Throughput [\SI{}{inference/\second}]}\\
Kooi and Babu\v{s}ka~\cite{inclined_landing} & STM32F405 & RL for landing & R & 250* & \SI{4.7}{\kilo\nothing} & \SI{4.7}{\kilo\nothing} & 400\\
Shi~\textit{et al.~}~\cite{neural_swarm_2} & STM32F405 & Interaction forces & R & 250* & \SI{9.5}{\kilo\nothing} & \SI{9.5}{\kilo\nothing} & —\\
% Cereda \textit{et al.}~\cite{cereda2021improving} & GWT GAP8 & Pose estimation & R & 96 & \SI{303}{\kilo\nothing} & \SI{14.7}{\mega\nothing} & 48~\\
Cereda~\textit{et al.}~\cite{cereda_nas} & GWT GAP8 & Pose estimation & R & 90 & \SI{65}{\kilo\nothing} & \SI{7.4}{\mega\nothing} & 51\\
% Bouwmeester \textit{et al.}~\cite{decroon_nanoflownet} & GWT GAP8 & Navigation & S & --- & \SI{171}{\kilo\nothing} & --- & 5.6\\
Lamberti \textit{et al.}~\cite{lamberti_exploration_detection} & GWT GAP8 & Object detection & C+R & 134 & \SI{4.7}{\mega\nothing} & \SI{534}{\mega\nothing} & 1.6\\
Niculescu \textit{et al.}~\cite{pulpdronetv2JETCAS} & GWT GAP8 & Navigation & C+R & 102 & \SI{320}{\kilo\nothing} & \SI{41}{\mega\nothing} & 19\\
\textbf{Ours} & \textbf{GWT GAP8} & \textbf{Navigation} & \textbf{C+R} & \textbf{100} & \textbf{\SI{2.9}{\kilo\nothing}} & \textbf{\SI{1.1}{\mega\nothing}} & \textbf{139}
\end{tblr}
\begin{tablenotes}
  \small
  \item *estimated
\end{tablenotes}
\end{threeparttable}
\end{table*}

\begin{figure}[t]
\centering
\includegraphics[width=\linewidth]{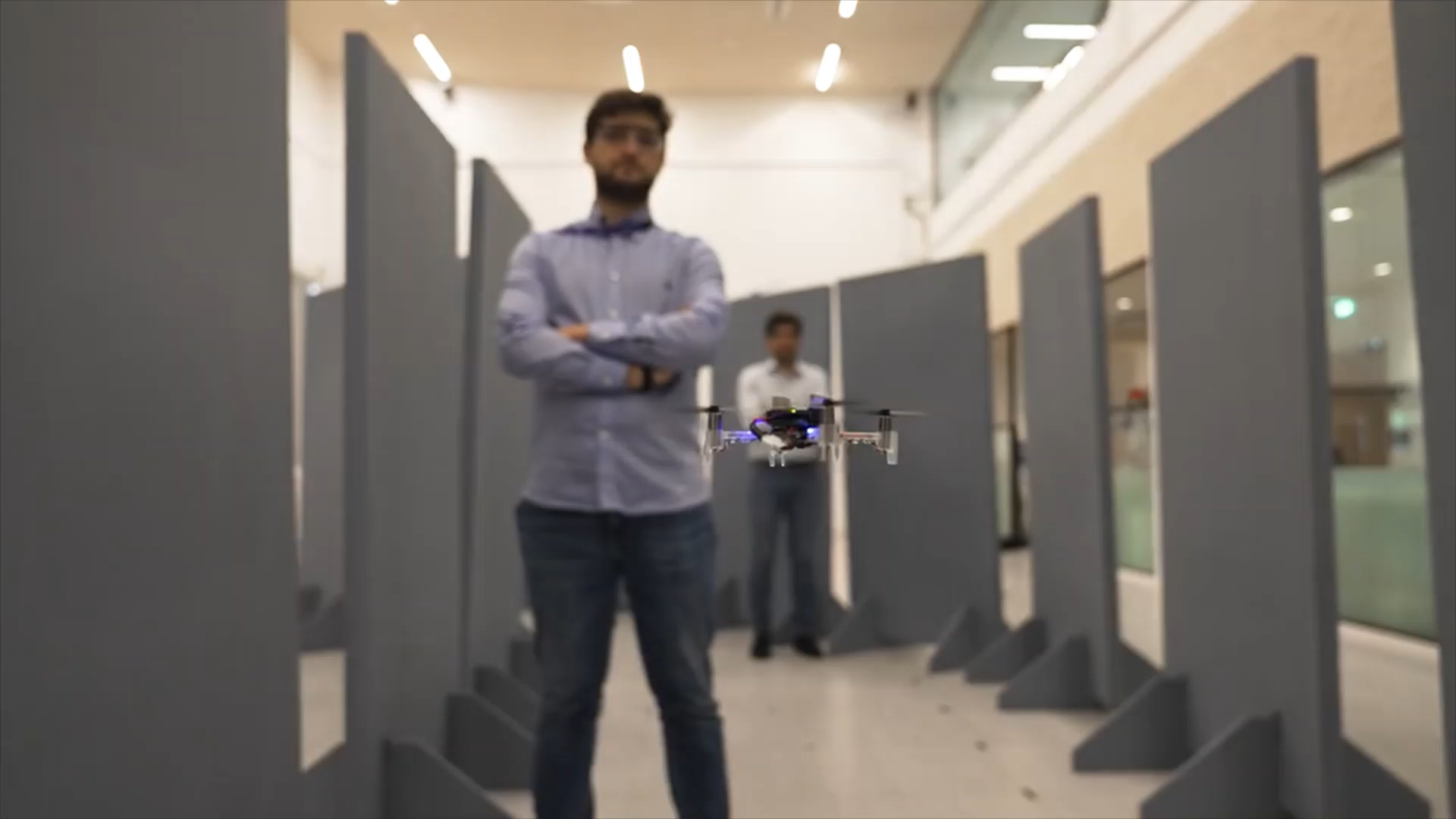}
\caption{Our autonomous nano-UAV navigating an unknown environment.}
\label{fig:intro}
\end{figure}

\IEEEPARstart{W}{ith} the growth of the Internet of Things (IoT), autonomous nano-sized unmanned aerial vehicles (UAVs) empowered with onboard artificial intelligence (AI) are becoming increasingly important~\cite{uavs_iot_survey}.
With a diameter of less than \SI{10}{\centi\meter} and weighing only tens of grams, these nano-UAVs can serve as ubiquitous IoT nodes, autonomously navigating environments while sensing and analyzing their surroundings~\cite{uav_iot_sensing}.
Their compact form factor allows them to operate in narrow/cluttered spaces~\cite{higgins_pathplanning_cluttered,pulpdronetv2JETCAS} and safely in the proximity of humans~\cite{cereda2021improving} (as shown in Figure~\ref{fig:intro}).
These characteristics enable nano-UAVs to be suitable in many use cases, such as exploring hazardous indoor environments~\cite{uav_chemicals} and in rescue missions~\cite{uav_iot_survey, uav_survey_applications}.

To act as autonomous smart IoT sensors, nano-UAVs must execute concurrent real-time tasks entirely onboard, including multiple heavy AI perception workloads~\cite{pulpdronetv2JETCAS, cereda2021improving, lamberti_exploration_detection, zhang2024_depthestimation_slam, drone_racing_survey}. 
However, achieving such a high level of autonomy presents significant challenges. 
Their extremely limited payload restricts them to accommodate only ultra-low-power microcontroller units (MCUs) that have stringent computational and memory constraints, e.g., a few hundred \SI{}{\kilo\byte} of on-chip memory and few giga operations per second (\SI{}{\giga Ops/\second}).
These constraints have prevented the deployment of multiple AI tasks on nano-UAVs.
Therefore, overcoming the computational/memory burden imposed by MCUs becomes paramount. 
We focus on optimizing and minimizing the AI workloads without compromising the drone’s behavior when stressed in real-world testing scenarios.

We specifically address a critical AI task required by any IoT node moving within an environment: \textit{visual navigation}, i.e., the capability to autonomously navigate an environment by relying solely on local visual information while avoiding collisions with obstacles.
The State-of-the-Art (SoA) convolutional neural network (CNN) for visual-based autonomous navigation on nano-UAVs is PULP-Dronet v2~\cite{pulpdronetv2JETCAS}, outputting a collision probability and a steering angle following visual cues in an environment, such as lanes.
This CNN has been deployed on a nano-sized UAV and demonstrated in both indoor and outdoor environments, enabling turns and avoiding collisions with dynamic obstacles.

However, this CNN has some shortcomings.
First, its navigation capabilities come at a non-negligible computational cost, allowing for a maximum throughput of \SI{19}{\fps} when running on the SoA Greenwaves Technologies (GWT) GAP8 System-on-Chip (SoC)~\cite{pulpdronetv2JETCAS}.

Moreover, we show that this CNN performs poorly when navigating static obstacles, limiting its real-world applicability. 
This limitation stems from the training dataset featuring disjoint image sets for steering and collision labels~\cite{pulpdronetv2JETCAS}.
A solution to this issue is to collect new data that integrates joint information on obstacle presence and yaw rate into all the training images.

In this paper, we set a new milestone in the SoA of autonomous visual-based nano-UAV navigation by shrinking the CNN memory footprint and computational burden while simultaneously improving its ability in static obstacle avoidance.
Our key contributions are:

\begin{itemize}
    \item we develop a methodology for dataset collection on a nano-UAV. 
    We collect \textit{unified collision avoidance and steering} information only with nano-UAV onboard resources, without dependence on external infrastructure like motion capture systems. 
    % (Section \ref{sec:dataset_methodology}). 
    The resulting \textit{PULP-Dronet v3} dataset consists of \SI{66}{\kilo\nothing} labeled images, which we release open-source along with our dataset collection framework;
    \item we train and deploy a novel family of CNNs, i.e., PULP-Dronet v3, that enables \textit{visual-based static obstacle avoidance and autonomous navigation} on nano-UAVs; 
    %(Section \ref{sec:cnn_architectures}, \ref{subsec:architecture_choice});
    \item we perform an extensive study for minimizing the memory footprint and computational complexity of CNNs for autonomous nano-UAV navigation.
    The resulting CNNs offer different trade-offs between in-field performance and model size.
\end{itemize}

We validate all our CNNs on our collected dataset and characterize their end-to-end execution time on the GAP8 SoC.
Our tiniest CNN, i.e., Tiny-PULP-Dronet v3, has only \SI{2.9}{\kilo\nothing} parameters and \SI{1.1}{\mega\nothing} multiply-accumulate (MAC) operations, $168\times$ smaller and $7.3\times$ faster (up to \SI{139}{\fps}) than the baseline CNN, i.e, the SoA PULP-Dronet v2~\cite{pulpdronetv2JETCAS}, when running on the same PULP GAP8 SoC.
This outcome allows us to free up precious computational resources that can be exploited to tackle additional tasks onboard.
This CNN only requires \SI{0.7}{\milli\joule} for each inference, with an average power consumption of \SI{100}{\milli\watt} when running on the GAP8's cores at their maximum frequencies.

Furthermore, we deployed and extensively tested in-field two CNNs: a larger one (PULP-Dronet v3), matching the size of the baseline CNN, and our tiniest distilled CNN (Tiny-PULP-Dronet v3). 
These tests evaluate how reducing the computational CNN workload impacts the navigation performance. 
In a challenging scenario with a narrow corridor containing four static obstacles and a \SI{180}{\degree} turn, both CNNs outperformed the baseline one, as shown in the supplementary video.
At a target speed of \SI{0.5}{\meter/\second}, our large and tiny CNNs successfully navigate through the corridor with a $80\%$ and $100\%$ success rate, respectively, while the baseline CNN consistently fails.
At higher speeds, the larger CNN demonstrated better robustness, succeeding 60\% of the time at a target speed of \SI{1}{\meter/\second}, whereas the tiny CNN failed.
We foster the research on autonomous nano-UAV navigation by releasing our new dataset, dataset collection framework, CNN weights, and code as open-source.

The remainder of this work is organized as follows.
Section~\ref{sec:related_works} provides an overview of tinyML for autonomous nano-UAVs.
Section~\ref{sec:background} covers the background of our work.
Section~\ref{sec:dataset_methodology} details our dataset and collection methodology.
Section~\ref{sec:cnn_architectures} describes our CNN design/shrinking approach.
Section~\ref{sec:results} compares the CNNs we introduced with experimental results.
Section~\ref{sec:infield_testing} describes the in-field experiments with our nano-drone.
Finally, Section~\ref{sec:conclusion} concludes the paper. 

\section{Related Work} \label{sec:related_works}

\subsection{TinyML on autonomous nano-sized UAVs}\label{sec:related_works_uavs}

This section presents an overview of lightweight and compact deep learning (DL) algorithms deployed on autonomous nano-UAVs, specifically nano-drones, which we summarize in Table~\ref{tab:rw}.
To cope with the limited resources of nano-UAVs, multiple works in literature aim to achieve only minimal functionalities and/or rely on low-dimensional input signals (e.g., no images) to minimize execution workload~\cite{inclined_landing, neural_swarm_2, inverted_landing}.

In the work of Kooi and Babu\v{s}ka~\cite{inclined_landing}, a deep reinforcement learning method employing proximal policy optimization enables autonomous landings of nano-drones on inclined surfaces. 
The CNN designed for this purpose requires around \SI{4.5}{\kilo\mac} operations per forward step. 
Although the CNN operates efficiently, computing an inference in approximately \SI{2.5}{\milli \second} on a single-core Cortex-M4 processor, its functionality remains constrained solely to the landing task.
Similarly utilizing the Cortex-M4 processor, Neural-Swarm2~\cite{neural_swarm_2} leverages a controller based on DL for managing interaction forces during nano-drones' formation flights. 
With a processing cost of $\sim$\SI{9.5}{\kilo \mac}, each nano-drone in the swarm processes the relative position and velocity data of surrounding UAVs, tackling safe maneuvers in close proximity with other UAVs.

To overcome computational constraints posed by single-core MCUs, multi-core flight controllers designed for AI workloads address the limitations imposed by higher-complexity DL-based workloads.
The SoA MCU for commercial off-the-shelf (COTS) nano-UAVs is the GAP8 SoC, an embodiment of the parallel ultra-low-power (PULP) paradigm with a general-purpose 8-core parallel cluster.
This SoC offers a peak throughput of \SI{5.4}{\giga Ops/\second}~\cite{garofalo2020pulpnn} and \SI{512}{\kilo\byte} of on-chip memory.
This fully programmable MCU was successfully exploited to enable the execution of more sophisticated visual-based AI workloads~\cite{cereda2021improving, lamberti_exploration_detection, pulpdronetv2JETCAS}.

Cereda \textit{et al.}~\cite{cereda2021improving} exploited GAP8 to demonstrate a fully autonomous nano-drone performing a human pose estimation task.
Their CNN, called PULP-Frontnet, predicts the drone's relative pose with respect to a freely moving human subject. 
This prediction aims at maintaining a consistent distance in front of human subjects while following their movements, and it only exploits low-resolution grayscale images captured from a front-facing camera.
Their model requires \SI{14.7}{\mega \mac} and achieves an inference rate of \SI{48}{\fps} while consuming \SI{96}{\milli\watt}.
\cite{cereda_nas} exploited neural architecture search techniques to design a tinier version of the PULP-Frontnet CNN, obtaining a \SI{7.4}{\mega \mac} and \SI{65}{\kilo \byte} model, while retaining good regression performance when tested in-field.
Like our work, their approach to CNN architecture design explores structures inspired by Mobilenet v2~\cite{mobilenetv2}.

Exploiting the same GAP8 SoC, Lamberti \textit{et al.}~\cite{lamberti_exploration_detection} tackled a more complex high-level task for nano-UAVs: object detection, which has a computational complexity significantly greater (over an order of magnitude higher) than~\cite{cereda2021improving, cereda_nas}.
They deployed a \SI{4.7}{\mega\byte} CNN capable of detecting two classes of objects, demonstrating its application in the context of an exploration and search mission.
Such workload requires a significant effort of \SI{534}{\mega \mac} per inference and exceeds the SoC on chip L2 memory of \SI{512}{\kilo\byte}, introducing additional latency cost for transferring data between on-chip and off-chip memories. 
As a result, the system only achieved a throughput of \SI{1.6}{\fps} when tested onboard.

Zhang \textit{et al.}~\cite{zhang2024_depthestimation_slam} deployed a compute-intensive CNN with \SI{310}{\kilo\nothing} parameters for relative visual-based depth estimation on a nano-drone. 
The depth information from the front-facing grayscale camera is exploited for obstacle avoidance.
However, the CNN runs onboard only at \SI{0.2}{\fps} and \SI{1.2}{\fps} on QVGA and QQVGA images, respectively, limiting the navigation capabilities of the drone to low speeds.
Niculescu \textit{et al.}~\cite{pulpdronetv2JETCAS} used instead a CNN for enabling end-to-end visual-based autonomous navigation on a nano-drone embedding the GAP8 SoC.
The CNN, namely PULP-Dronet v2, has a computational cost of $\sim$\SI{41}{\mega \mac} per inference and requires \SI{320}{\kilo \byte} of memory footprint.
When tested on an autonomous nano-drone, this neural network allowed it to tackle turns and avoid dynamic obstacles appearing along the way.

However, PULP-Dronet v2 has two main limitations.
First, as highlighted by the authors~\cite{pulpdronetv2JETCAS}, this CNN lacks the ability to guide the nano-UAV around static obstacles.
This poses a significant limitation in its practical application within challenging real-world scenarios.
This missing capability is attributed to the training dataset of PULP-Dronet v2, as it comprehends multiple sets of images with disjointed labels for the steering and collision tasks~\cite{pulpdronetv2JETCAS}.
The second limitation of this work is its non-negligible workload of $\sim$\SI{41}{\mega \mac}, limiting its throughput to a maximum of \SI{19}{\fps}. 
Despite being enough to enable autonomous navigation, this system would greatly benefit from solving the same task with only a minimal fraction of the current workload, as it would free precious resources that can be exploited to tackle additional tasks~\cite{tiny_dronets, uav_iot_survey, uav_survey_applications}.
\cite{tiny_dronets} demonstrates that the PULP-Dronet v2 CNN architecture is over-parametrized for the task it solves. 
However, it lacks an in-field demonstration that a smaller neural network can achieve sufficient accuracy for safe flight.

Therefore, in our work we move from PULP-Dronet v2, focusing on: \textit{i}) enabling new navigation skills, i.e., tackling the static obstacle avoidance scenario, demonstrating with in-field results that the dataset was the limiting factor for achieving this task. \textit{ii}) we enable autonomous navigation with a CNN that is $168\times$ smaller than PULP-Dronet v2, leaving plenty of memory and computing resources that can be exploited to embed additional AI tasks on the nano-drone.

\subsection{Datasets for nano-UAV autonomous navigation}

As we highlighted the importance of introducing a new dataset for visual-based autonomous nano-drone navigation in Section~\ref{sec:related_works_uavs}, we review the datasets for UAVs in the literature.
Dupeyroux \textit{et al.}~\cite{decroon_dataset} released a dataset for obstacle detection and avoidance, specifically targeting micro-sized drones (i.e., weighting $\sim$\SI{0.5}{\kilo\gram}). 
It has \SI{92}{\giga\byte} of data, including full-HD images for $\sim80\%$ of the acquisitions, and the drone pose is tracked with a motion capture system.
Thus, adapting these images to the nano-drone use case would require a photometric augmentation pipeline to convert full-HD images to the format of a low-quality and low-resolution camera commonly found on nano-UAVs~\cite{pulpdronetv2JETCAS}.
Conversely, we collect a dataset exploiting nano-drones exclusively: we log the low-quality images from the onboard camera along with the pilot's yaw-rate input, which is fed to the flight controller during the data acquisition.
This approach allows us to train an end-to-end CNN for autonomous navigation by exploiting the pilot's input commands rather than the UAV's external tracking.

The Dronet dataset, introduced by Loquercio \textit{et al.}~\cite{loquercio2018dronet}, was created to enable autonomous navigation on large UAVs.
They merged two distinct image sets: one with \SI{39.1}{\kilo\nothing} high-resolution images from car driving scenarios labeled only with steering information, and another with \SI{32.2}{\kilo\nothing} high-resolution images from bicycle rides labeled with binary collision information. 
This dataset was expanded with approximately  $\sim1300$ grayscale low-quality images collected from a nano-UAV and labeled only with binary collision information~\cite{pulpdronetv2JETCAS}. 
The combination of these three disjointed sets of images forms the Himax+Original dataset used to train PULP-Dronet, which we exploit for comparison in Section~\ref{sec:results}.

However, this dataset does not provide joint steering and collision labels for the navigation, leading to poor performance when tackling static obstacle avoidance~\cite{pulpdronetv2JETCAS}, as detailed in Section~\ref{sec:related_works_uavs}.
Similarly, ~\cite{drone_dataset_hdin} collected $\sim$\SI{15}{\kilo\nothing} images for solving the same task as Dronet~\cite{loquercio2018dronet}.
However, this dataset is also divided into two disjointed sets of images for steering and collision labels.
~\cite{navardi_pulp_dronet} used the Dronet dataset to train an off-board CNN for nano-drones autonomous navigation, still not tackling the static obstacle avoidance scenario.

The dataset presented in this paper overcomes these limitations as all the $\sim$\SI{66}{\kilo\nothing} collected images feature both collision and steering information, i.e., it does not have disjointed labels.
Specifically, we log the input of a human pilot navigating a nano-drone in multiple environments.
These labels can be used to train an end-to-end CNN that imitates the behavior of a human pilot.
Furthermore, we log the drone's estimated state and label all the images with a binary label for obstacle avoidance.
We open-source our dataset, dataset collector framework, and pre-trained models to foster research on autonomous nano-drone navigation.

\section{Nano-UAV robotic platform} \label{sec:background}

\begin{figure*}[t]
\centering
\includegraphics[width=\linewidth]{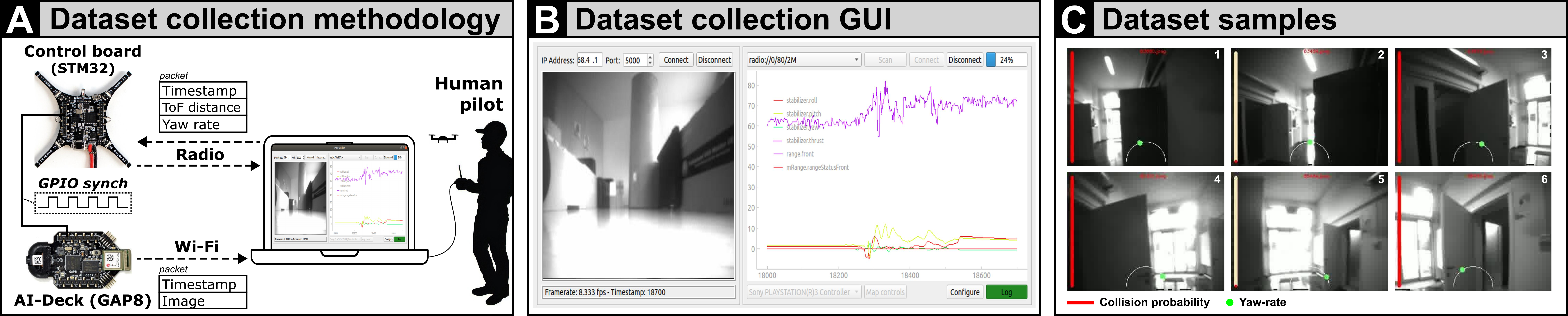}
\caption{A) our dataset collection methodology, B) our dataset collector GUI, and C) a sample sequence from our collected dataset.}
\label{fig:methodology}
\end{figure*}

\subsection{Robotic platform} \label{sec:background_robotic_platform}

The robotic platform employed in this work encompasses the Bitcraze Crazyflie 2.1 quadrotor, a commercially available nano-drone weighing \SI{27}{\gram}  and measuring 
\SI{10}{\centi\meter} in diameter. 
This open-source and open-hardware drone uses the STM32F405 MCU as its flight controller coupled with the Bosch BMI088 inertial measurement unit (IMU), which combines an accelerometer and gyroscope.
The STM32 MCU operates at speeds up to \SI{168}{\mega\hertz} and integrates \SI{48}{\kilo\byte} static RAM (SRAM) and \SI{128}{\kilo\byte} flash memories, facilitating onboard inertial state estimation and actuation control.
The IMU data drives an extended Kalman filter for state estimation at a rate of \SI{100}{\hertz}, while a proportional-integral-derivative control loop cascade manages actuation. 
This cascade comprises two control loops, with one governing attitude at \SI{500}{\hertz} and the second updating position at \SI{100}{\hertz}.
The Crazyflie 2.1 also integrates a nRF51822 MCU for \SI{2.4}{\giga\hertz} radio band communication, which we use only for the purpose of dataset collection (Section~\ref{sec:dataset}).

Our configuration extends the robotics platform with two pluggable printed circuit boards (PCBs) from Bitcraze: the Flow deck and AI-deck. 
The \SI{3.5}{\gram} Flow deck incorporates the PMW3901 optical flow visual sensor and the VL53L1x time-of-flight (ToF) ranging module. 
The optical flow sensor detects drone motion in multiple directions, while the ToF sensor measures distance from the ground. These inputs enhance onboard state estimation accuracy, minimizing long-term drift.

The second expansion board, the AI-deck, serves as the primary onboard computing unit responsible for executing heavy AI workloads,  such as the proposed autonomous navigation CNNs. 
This PCB, weighing \SI{4.4}{\gram}, integrates the GWT GAP8 SoC as the onboard computer for visual processing and AI-driven autonomous navigation.
GAP8 is accompanied by onboard memories: \SI{64}{\mega\byte} of  HyperFlash and \SI{8}{\mega\byte}  of HyperRAM. 
The PCB also mounts a Himax HM01B0 ultra-low-power monochrome QVGA camera, an ESP32-based Wi-Fi transceiver, and a UART communication channel between the STM32 and the GAP8. 
In this work, we focus on a configuration where the power-intensive Wi-Fi remains off, aiming for a fully autonomous system where all navigation intelligence resides onboard the nano-drone without relying on external communication or computation.
However, we exploited the high bandwidth of the Wi-Fi communication for the dataset collection, as detailed in Section~\ref{sec:dataset}.

For the sake of dataset collection only (Section~\ref{sec:dataset}), we add an extra PCB to our drone, i.e., the \textit{Multi-ranger deck}.
This deck weighs \SI{2.3}{\gram} and comprises five single-beam VL53L1x ToF distance sensors placed on the drone's top and lateral sides.
These sensors provide line-of-sight distance measurements with a frequency of \SI{20}{\hertz} and have an operative range that goes from \SI{0}{\meter} to \SI{4}{\meter}.

\subsection{GAP8 System-on-Chip} \label{sec:background_gap8}

The GWT GAP8 SoC\footnote{\url{https://greenwaves-technologies.com/low-power-processor/}}, which is used in this work for all onboard vision and tinyML processing, is an ultra-low-power SoC that combines the capabilities of a regular MCU, the \textit{Fabric Controller} (FC) domain, with a fully programmable parallel accelerator called \textit{Cluster} (CL) that can yield up to \SI{5.4}{\giga Ops/\second}~\cite{garofalo2020pulpnn} in a power envelope of $\sim$\SI{100}{\milli\watt}.
The FC domain includes a low-power processor implementing the RISC-V \texttt{RV32IMCXpulpv2} instruction set architecture (ISA) based on the RI5CY design, paired with on-chip SRAM memory and peripherals supporting protocols like SPI, I2C, HyperBus, and Camera Parallel Interface (CPI).
These can be accessed through a dedicated Direct Memory Access (DMA) engine called $\mu$DMA to offload the communication burden from the FC.

The CL accelerates heavier parallelizable workloads typical of digital signal processing and AI applications.
It includes 8 RISC-V cores with the same ISA extensions as the FC; in particular, the \texttt{Xpulpv2} extension includes 8-bit and 16-bit packed single instruction multiple data (SIMD), multiply-accumulate, and dot-product operations, which enhance the SoC's linear algebra capabilities for such workloads.
The cores are connected over a logarithmic interconnect to \SI{64}{\kilo\byte} of L1 Tightly Coupled Data Memory (TCDM) comprising 16 memory banks.
The logarithmic interconnect assures 1-cycle latency access for all the cores when there is no bank conflict, enabling fast data parallelism among the cores.
The cluster has a DMA to offload data transfers between the TCDM and the \SI{512}{\kilo\byte} L2 memory.
The cores are programmed with the Single-Program Multiple-Data programming model and synchronized using dedicated hardware for low-latency barriers.

The FC and cluster domains are separately clocked to tune for best energy efficiency and performance trade-off.
The FC domain can be clocked between \SI{50}{\mega\hertz} and \SI{250}{\mega\hertz}, while the cluster domain can be between \SI{100}{\mega\hertz} and \SI{175}{\mega\hertz}.
The SoC voltage ($V_{dd}$) can be set between \SI{1}{\volt} and \SI{1.2}{\volt} depending on the FC and cluster clock frequency.

\subsection{Automatic Deployment Tools} \label{sec:background_deployment}

\begin{figure*}[t]
\centering
\includegraphics[width=1\linewidth]{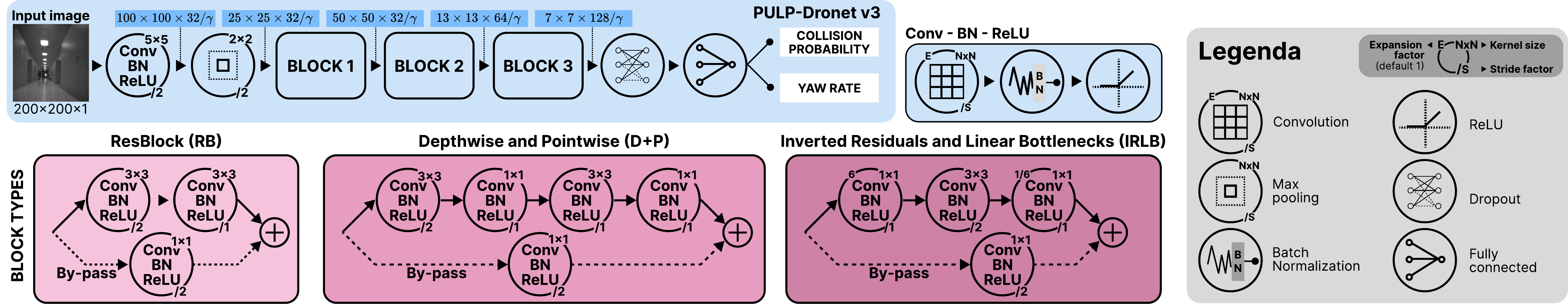}
\caption{Our CNN architecture exploration includes: \textit{i}) three block types -- RB, D+P, IRLB; \textit{ii}) an optional bypass connection (dashed line); \textit{iii}) variations on the number of channels based on $\gamma$. \blue{Output feature map sizes are  represented as ($Width\times Height \times  Channels$).}}
\label{fig:cnn_architecture}
\end{figure*}

Developing CNNs for an MCU-class processing device, like the GAP8 on our nano-drone, is a multi-objective optimization problem.
In our case, it must take into account: \textit{i}) memory limitations (L2 {\SI{512}{\kilo\byte} and L1 {\SI{64}{\kilo\byte}}), \textit{ii}) the hardware limitations (i.e., no floating point unit), power envelope ($\sim$\SI{100}{\milli\watt}), and throughput (i.e., a flying drone needs to react in real-time).
As for PULP-Dronet v2~\cite{pulpdronetv2JETCAS}, we use an automated flow that works in two steps: first, we quantize the neural network~\cite{quantization_survey}, and then we perform a hardware-aware deployment of the quantized model.

For quantization, we take \texttt{float32} pre-trained CNN models, and we apply post-training quantization using NEMO~\cite{conti2020nemo}, a deep neural network (DNN) quantization tool that performs uniform asymmetric static layer-wise quantization.
We apply fixed-point 8-bit (\texttt{int8}) post-training quantization to the weights and activations of our CNNs.
By doing so, we enable optimized 8-bit fixed-point arithmetic on GAP8, i.e., packed-SIMD instructions~\cite{garofalo2020pulpnn}.
Moreover, the conversion from \texttt{float32} data type and the quantized \texttt{int8} reduces by $4\times$ the memory footprint of the CNNs models.

For hardware-aware deployment, we use DORY\cite{burrello2020dory}, a SoA code generation tool for quantized DNNs.
It is tuned for deployment on memory-constrained embedded devices with multiple levels of memory hierarchy to optimize performance.
To fit the data in the available memory resources, large operations need to be split into smaller pieces called tiles.
DORY models the tile size optimization problem as a constraint
programming problem with the memory size as the main constraint and hardware-aware heuristics to guide the ILP solver toward the best-performing solutions.

While tiling makes the operations fit into our desired memory, it still requires data marshaling between the memory levels to process the whole layer.
DORY overlaps this data movement cost with computation by employing multi-buffering and software pipelining to hide this data movement cost.
For efficient operation computation, DORY exploits the PULP-NN~\cite{garofalo2020pulpnn} open-source library, which offers hand-optimized kernels for quantized neural networks executing on the PULP cluster.

\subsection{The baseline PULP-Dronet v2 CNN} \label{subsec:background_pulp_dronet}

PULP-Dronet v2~\cite{pulpdronetv2JETCAS} is an end-to-end vision-based CNN for autonomous navigation aboard nano-drones and suited for deployment on the AI-deck.
We employ PULP-Dronet v2 as a baseline model for comparison with our work.
The architecture of this CNN is represented in Figure~\ref{fig:cnn_architecture}, and it is based on a sequence of three residual blocks~\cite{resnet2022}, which we denote as ResBlock (RB)}.
Each block consists of two parallel branches: the main branch performs two $3\times3$ convolutions, while the by-pass one performs one $1\times1$ convolution.
The number of output channels is 32, 64, and 128 for the three blocks, respectively.
For each block, the last convolution on both branches applies a stride factor of 2 to half the feature map size.
Following each convolution is a Batch Normalization (BN) layer and a rectified linear unit (ReLU) activation function.
The ReLU output is capped to the value 6, therefore called ReLU6.

The CNN processes a grayscale image of size $200\times200$. 
This image is the bottom-center crop extracted from the QVGA image captured by the AI-deck's Himax camera.
The final layer produces two distinct outputs: a collision probability (classification problem) and a steering angle (regression problem). 
Consequently, the model is trained using two distinct metrics: the mean squared error (MSE) and the binary cross-entropy (BCE). 
These metrics are combined into a unified loss function, $Loss = MSE + \beta BCE$, which we also employ in this work.  
$\beta$ is set to 0 for the first 10 epochs, gradually increasing in a logarithmic way to prioritize the regression problem.
The training employs a dynamic \textit{negative hard-mining} approach, gradually focusing the loss computation on the k-top samples exhibiting the highest error.
PULP-Dronet v2 is deployed in fixed-point 8-bit arithmetic on the multi-core GAP8 SoC, yielding \SI{41}{\mega \mac} operations per frame and \SI{320}{\kilo\byte} of weights.

\begin{figure}[t]
\centering
\includegraphics[width=1\linewidth]{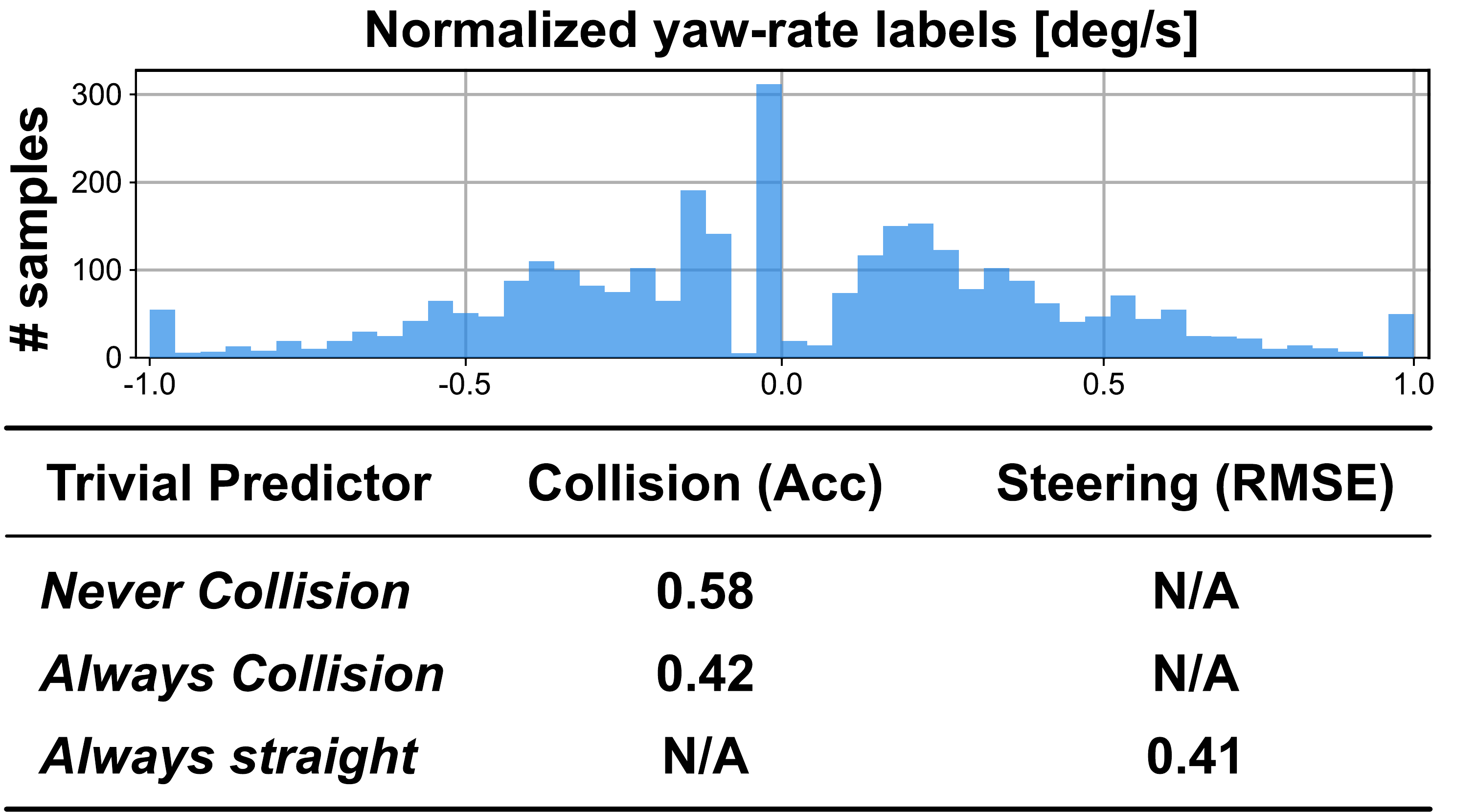}
\caption{Distribution of the yaw-rate labels (normalized in [-1,+1]) of our testing dataset and classification/regression performances of three trivial predictors.}
\label{fig:trivial_distribution}
\end{figure}

\section{Dataset collection methodology} \label{sec:dataset_methodology}

\subsection{Dataset collection framework} \label{sec:dataset_collector_framework}

We developed a software framework for dataset collection, outlined in Figure~\ref{fig:methodology}-A.
It enables a pilot to manually control the nano-drone and simultaneously log:
\textit{i}) any data from the STM32 flight controller (e.g., the drone's state estimation) and sensors attached to it (e.g., additional expansions decks of the Crazyflie 2.1\footnote{\url{https://store.bitcraze.io/collections/decks}}, such as ranging sensors), 
\textit{ii}) images captured by the AI-deck's front-facing camera.
All data is recorded on a PC base station, which independently receives the two data streams: the STM32 packets, transmitted via radio, and AI-deck images, transmitted via Wi-Fi.

To ensure the data from these two separated streams can be accurately matched in post-processing, we \textit{i}) 
periodically synchronize both the STM32 and GAP8 MCUs with a global clock, generated from a GPIO of GAP8, and \textit{ii}) send each chuck of data associated with a timestamp that is globally synchronized across our robotic platform.
This synchronization methodology minimizes matching errors by ensuring each packet is timestamped at its source, thus avoiding errors related to wireless communication and leaving only minimal discrepancies caused by the MCUs' oscillator drifts.

Our dataset collector framework consists of: \textit{i}) code for the STM32 and GAP8 SoC (Figure~\ref{fig:methodology}-A), which enables sending the data streams while keeping the global clock synchronized, and \textit{ii}) a graphical user interface (GUI) (Figure~\ref{fig:methodology}-B) to plot the data streamed from the flight controller and visualize images collected from the AI-deck.

\subsection{Our dataset for nano-drone autonomous navigation} \label{sec:dataset}

We introduce a new dataset for visual-based autonomous nano-drone navigation.
We move on from the limitations of the past PULP-Dronet dataset~\cite{pulpdronetv2JETCAS}, which was created by assembling different sets of data, each one having images labeled either for the steering or the collision avoidance task, as explained in~\ref{sec:related_works_uavs}, ultimately penalizing the static obstacle avoidance~\cite{pulpdronetv2JETCAS}.
To tackle this limitation, we collect a new unified dataset from scratch. 

With the dataset collector framework introduced in Section~\ref{sec:dataset_collector_framework}, we collected a dataset of \SI{66}{\kilo\nothing} images for nano-drones' autonomous navigation, for a total of $\sim$\SI{600}{\mega\byte} of data.
We used the Bitcraze Crazyflie 2.1, described in Section~\ref{sec:background_robotic_platform}.
A human pilot manually flew the drone, collecting \textit{i}) images from the grayscale QVGA Himax camera sensor of the AI-deck, \textit{ii}) the gamepad's yaw-rate, normalized in the $[-1;+1]$ range, inputted from the human pilot, \textit{iii}) the drone's estimated state,  and \textit{iv}) the distance between obstacles and the drone measured by the front-looking ToF sensor.

After the data collection, we labeled all the images with a binary collision label whenever an obstacle was in the line of sight and closer than \SI{2}{\meter}.
We recorded 301 sequences in 20 different environments. 
Each sequence of data is labeled with high-level characteristics: scenario (i.e., indoor or outdoor), path type (i.e., presence or absence of turns), obstacle types (e.g., pedestrians, chairs),  flight height (i.e., \SI{0.5}{\meter}, \SI{1}{\meter}, \SI{1.5}{\meter}), light conditions (dark, normal, bright), acquisition date, and a location name identifier.

For training and validating our CNNs, we post-processed the datasets as follows.
We used $70\%$, $10\%$, and $20\%$ of the images as training, validation, and testing sets, respectively.
We augmented the training images by applying random flipping, brightness augmentation, vignetting, and blur. 
The resulting training dataset has $\sim$\SI{124}{\kilo\nothing} images, split as follows: \SI{109}{\kilo\nothing}, \SI{5}{\kilo\nothing}, \SI{10}{\kilo\nothing} images for training, validation, and testing, respectively.
To address the labels' bias towards the center of the $[-1;+1]$ yaw-rate range in our testing dataset -- specifically, the over-representation of images associated with a yaw-rate of 0, indicating no input from the human pilot and thus the drone flying straight -- we balanced the dataset by selectively removing a portion of images that had a yaw-rate of 0.

The final distribution of our balanced dataset is represented in Figure~\ref{fig:trivial_distribution}.
In the same Figure, we also report the RMSE and Accuracy for three trivial predictors, i.e., a predictor that either always predicts collision to 1 or 0, and a predictor that always predicts a yaw-rate of zero (i.e., go straight).
These values can be later used in Section~\ref{sec:results} as a baseline to assess the RMSE and Accuracy performance of our trained CNNs.

\section{Neural network architecture design} \label{sec:cnn_architectures}

The original PULP-Dronet v2 CNN's architecture~\cite{pulpdronetv2JETCAS} exploits three ResBlocks (RB)~\cite{resnet2022}, as described in Section~\ref{subsec:background_pulp_dronet}.
In other vision tasks, the RB architecture has been largely replaced by other kinds of CNN tolopogies, which provide similar accuracy characteristics but with lower workload and memory footprint~\cite{cereda_nas}.
Therefore, we design the new PULP-Dronet v3 architecture by exploring several modifications to its baseline topology. 

Our modifications, detailed in Figure~\ref{fig:cnn_architecture}, involve substituting the three RB blocks of PULP-Dronet v2, removing parallel by-pass branches, and varying the number of channels in the intermediate feature maps.
To optimize autonomous navigation and enable more complex onboard functionality, we studied several modifications to this setup.
First, we replace these blocks with two other blocks we designed, taking inspiration from the well-known Mobilenet family of CNNs~\cite{mobilenetv1, mobilenetv2}.
These CNNs have been shown to be both accurate for visual AI tasks and suitable for efficient MCU deployment~\cite{cereda_nas}.
For each block we substitute, we keep the same output feature map dimensions (width and height) of PULP-Dronet v2.

\begin{table}
\centering
% \footnotesize
\caption{Cross-validation between PULP-Dronet v2 dataset and ours (v3).}
\label{tab:cross_validation}\begin{tblr}{
  row{odd} = {c},
  row{4} = {c},
  cell{1}{1} = {r},
  cell{1}{2} = {c=2}{},
  cell{1}{4} = {c=2}{},
  cell{2}{2} = {c},
  cell{2}{3} = {c},
  cell{2}{4} = {c},
  cell{2}{5} = {c},
  vline{2,4,6} = {2-4}{},
  hline{2-3,5} = {2-5}{},
}
\diagbox{\textbf{Training}}{\textbf{Testing}} & \textbf{v2} (\SI{6.9}{\kilo\nothing} images) &  & \textbf{v3} (\SI{10}{\kilo\nothing} images) & \\
 & \textbf{RMSE} & \textbf{Accuracy} & \textbf{RMSE} & \textbf{Accuracy}\\
\textbf{v2} (\SI{64}{\kilo\nothing} images) & 0.118 & 90\% & 0.556 & 54\%\\
\textbf{v3} (\SI{109}{\kilo\nothing} images) & 0.236 & 67\% & 0.350 & 83\%
\end{tblr}
\end{table}

The first new block we design is based on the Mobilenet v1~\cite{mobilenetv1}.
It uses separable depthwise and pointwise (D+P) convolutional layers instead of traditional ones. 
Such layers factorize a standard convolution into a sequence of K$\times$K \textit{depthwise} convolution and a $1\times1$ convolution called \textit{pointwise} convolution, where K is the kernel size.
We define our D+P block as consisting of two branches: the main branch performs two $3\times3$ D+P convolutions, while the parallel by-pass branch executes a standard $1 \times 1$ convolution. 
We apply convolutional strides different from 1 only to the last convolution in both branches.
Each convolution (depthwise, pointwise, and standard) is subsequently followed by a BN layer and a ReLU6. 
We refer to the sequence of these three layers as a Conv-BN-ReLU.
This Conv-BN-ReLU pattern simplifies both quantization and deployment explained in Section~\ref{sec:quantization_deployment}: the tensor outputted by the Conv operation demands a finer grain representation (utilizing \SI{32}{\bit}) compared to the inputs and weights.
Therefore, the BN is used to accumulate a \SI{32}{\bit} representation, while the ReLU reduces it back immediately to \SI{8}{\bit}.
 
The second block we define is inspired by Mobilenet v2~\cite{mobilenetv2}, using inverted residuals and linear bottlenecks (IRLB) layers.
This block comprises \textit{i}) a $1\times 1$ convolution, expanding the number of channels by an expansion factor (E), \textit{ii}) a D+P convolution, and \textit{iii}) a $1\times1$ projection convolution that inverts the expansion, reducing the number of output channels. 
Stride factors different from 1 are applied in the D+P convolution.
We choose an expansion ratio equal to 6 as in ~\cite{mobilenetv2}.
Additionally, we add a by-pass branch performing a $1\times1$ convolution, ensuring the same output size as the main IRLB branch.

We explore additional architecture variations by considering the removal of parallel by-pass branches from each block type.
These branches primarily serve to mitigate vanishing gradient effects in deep CNNs~\cite{resnet2022}. 
However, recent studies~\cite{tiny_dronets} have highlighted their inefficacy in shallow CNN models, such as the seven-layer PULP-Dronet.
Lastly, we vary the number of channels of the intermediate CNN feature maps to investigate the accuracy, memory footprint, and computational cost trade-offs.
As in ~\cite{tiny_dronets, mobilenetv1}, we thin the CNNs' tensors by applying a $\gamma$ dividing factor to the number of channels across all convolutional layers.
We span the $\gamma$ parameter in the range $[1, 2, 4, 8]$, where $\gamma=1$ corresponds to the baseline size for PULP-Dronet v2\cite{pulpdronetv2JETCAS}.

\begin{table}[tb]
\centering
\caption{\blue{Analysis of the PULP-Dronet v3 CNN architectures, varying its blocks and by-pass branches. The first row matches the SoA baseline architecture~\cite{pulpdronetv2JETCAS}.}}
\label{tab:architecture_choice_blocks}
\begin{tblr}{
  width = \linewidth,
  colspec = {Q[104]Q[170]Q[10]Q[108]Q[90]Q[69]Q[75]Q[82]Q[195]},
  cells = {c},
  hline{1,8} = {-}{0.08em},
  hline{2} = {-}{},
}
\textbf{Blocks} & \textbf{by-pass} & $\bm{\gamma}$ & \textbf{Type} & \textbf{RMSE} & \textbf{Acc} & \textbf{MAC} & \textbf{Param} & \textbf{Size [B]}\\
RB & yes & /1 & fp32 & 0.339 & 83\% & 41M & 320k & 1.3M\\
RB & no & /1 & fp32 & 0.339 & 83\% & 40M & 309k & 1.2M\\
D+P & yes & /1 & fp32 & 0.352 & 83\% & 14M & 63k & 252k\\
\textbf{D+P} & \textbf{no} & \textbf{/1} & \textbf{fp32} & \textbf{0.350} & \textbf{84\%} & \textbf{12M} & \textbf{51k} & \textbf{204k}\\
IRLB & yes & /1 & fp32 & 0.369 & 82\% & 43M & 140k & 560k\\
IRLB & no & /1 & fp32 & 0.364 & 83\% & 41M & 128k & 513k
\end{tblr}
\end{table}

\section{Experimental Results} \label{sec:results}

\subsection{Datasets evaluation} \label{subsec:dataset_comparison}

In Table~\ref{tab:cross_validation}, we assess the impact of training and testing on our new dataset (\textbf{v3}) versus the Himax+Original PULP-Dronet v2 dataset used in the SoA work~\cite{pulpdronetv2JETCAS} (\textbf{v2}); each dataset is split into different training and testing sets.
We perform this assessment by keeping the original PULP-Dronet v2 CNN architecture in both cases.
When the CNN is trained on v2 and tested on v2, we achieve performance consistent with the results reported in~\cite{pulpdronetv2JETCAS}.
When training on v3 and testing on v3, both performances decrease compared to the previous case due to the higher complexity of the v3 testing set, such as a more uniform distribution of steering labels and a richer, and therefore, more challenging, dataset.
We also report the cross-validation of training on v2 and testing on v3, and training on v3 and testing on v2.
In both cases, performances drop compared to training and testing on the same dataset version, but the v2-trained CNN (tested on v3) has a higher drop on both RMSE and Accuracy than the v3-trained (tested on v2), suggesting better generalization capabilities for the proposed dataset.

\subsection{CNN architectures exploration} \label{subsec:architecture_choice}

\begin{table}[tb]
\centering
\caption{\blue{Accuracy, RMSE, MACs, and memory footprint of our  CNN architectures by varying $\gamma$. The last row corresponds to Tiny-PULP-Dronet v3.}}
\label{tab:architecture_choice_size}
\begin{tblr}{
  width = \linewidth,
  colspec = {Q[104]Q[170]Q[10]Q[108]Q[90]Q[69]Q[75]Q[82]Q[195]},
  cells = {c},
  hline{1,6} = {-}{0.08em},
  hline{2} = {1-3,5-8}{},
  hline{2} = {4,9}{0.03em},
}
\textbf{Blocks} & \textbf{by-pass} & \bm{$\gamma$} & \textbf{Type} & \textbf{RMSE} & \textbf{Acc} & \textbf{MAC} & \textbf{Param} & \textbf{Size [B]}\\
D+P & no & /1 & fp32 & 0.350 & 84\% & 12M & 51k & 204k \\
D+P & no & /2 & fp32 & 0.367 & 84\% & 5.2M & 17k & 69k \\
D+P & no & /4 & fp32 & 0.373 & 81\% & 2.4M & 6.6k & 26k \\
D+P & no & /8 & fp32 & 0.379 & 78\% & 1.1M & 2.9k & 12k
\end{tblr}
\end{table}

To evaluate the PULP-Dronet v3 family of CNNs proposed in Section~\ref{sec:cnn_architectures}, we start by analyzing different block types (RB, D+P, and IRLB) and the effect of removing the by-pass branches.
In Table~\ref{tab:architecture_choice_blocks}, we assess the CNNs in terms of parameter count, MAC operations, and key performance metrics -- the Accuracy for the classification problem (the higher, the better) and the RMSE for the regression one (the lower, the better).

Removing the by-pass branches negligibly affects the CNN performance metrics in all cases, i.e., at most +0.005 RMSE and +1\% in accuracy with IRLB CNN.
On the other side, the by-pass removal brings the benefit of reducing the CNN size by $3-11\%$ and the number of MAC by $3-15\%$, depending on the architecture.
Focusing on the three CNNs without by-pass, the RB variant achieved the lowest RMSE of 0.339, whereas D+P and IRLB models score 0.350 and 0.369, respectively.
Instead, the D+P neural network achieves the highest performance of $84\%$ on the accuracy score.
The D+P CNN is also the faster CNN, requiring only \SI{12}{\mega \mac} per inference, and the smaller in memory usage, being \SI{6.2}{\nothing}$\times$ smaller than RB and \SI{2.5}{\nothing}$\times$ smaller than IRLB. 
Ultimately, we select the D+P model without by-pass branches for its efficiency and minimal performance drop.

\subsection{CNNs size analysis} \label{subsec:size_choice}

In this section, we analyze how the number of channels across our CNNs affects their memory requirements, number of operations, and regression/classification performances.
Starting from the CNN architecture D+P without by-passes, selected in Section~\ref{subsec:architecture_choice}, we span the $\gamma$ parameter (see Section~\ref{sec:cnn_architectures}) in the range $[1, 2, 4, 8]$, where $\gamma=1$ corresponds to the baseline size for PULP-Dronet v2\cite{pulpdronetv2JETCAS}.
Table~\ref{tab:architecture_choice_size} shows that using a dividing factor $\gamma=2$ does not affect the classification accuracy of the CNN when compared to $\gamma=1$, both scoring $84\%$.
When using smaller CNNs, the classification accuracy drops by $3\%$ with $\gamma=4$ and by $7\%$ with $\gamma=8$.

On the regression task, the RMSE gradually increases as the CNNs become smaller. 
For $\gamma=1$, the RMSE stands at \SI{0.350}{\nothing}, but increase to \SI{0.367}{\nothing}, \SI{0.373}{\nothing}, \SI{0.379}{\nothing} for $\gamma=2$, $\gamma=4$, $\gamma=8$, respectively.
On the other hand, the $\gamma$ factor greatly impacts the number of the CNN's parameters.
The biggest CNN ($\gamma=1$) has \SI{204}{\kilo\nothing} parameters, the CNN with ($\gamma=2$)  has $~3\times$ less parameters (\SI{69}{\kilo\nothing} parameters).
Increasing the $\gamma$ to 4 leads to a model with \SI{26}{\kilo\nothing} parameters, and finally the smallest CNN ($\gamma=8$), which we call Tiny-PULP-Dronet v3, leads to only  \SI{1.9}{\kilo\nothing} parameters.

\begin{table}[tb]
\centering
\caption{Accuracy, RMSE, and memory footprint of our quantized CNNs. \blue{The last row corresponds to Tiny-PULP-Dronet v3.}}
\label{tab:quantized_net}
\begin{tblr}{
  width = \linewidth,
  colspec = {Q[104]Q[170]Q[10]Q[108]Q[90]Q[69]Q[75]Q[82]Q[195]},
  cells = {c},
  hline{1,6} = {-}{0.08em},
  hline{2} = {1-3,5-8}{},
  hline{2} = {4,9}{0.03em},
}
\textbf{Blocks} & \textbf{by-pass} & \bm{$\gamma$} & \textbf{Type} & \textbf{RMSE} & \textbf{Acc} & \textbf{MAC} & \textbf{Param} & \textbf{Size [B]}\\
D+P & no & /1 & int8 & 0.361 & 84\% & 12M & 51k & 51k\\
D+P & no & /2 & int8 & 0.373 & 82\% & 5.2M & 17k & 17k\\
D+P & no & /4 & int8 & 0.378 & 81\% & 2.4M & 6.6k & 6.6k\\
D+P & no & /8 & int8 & 0.388 & 78\% & 1.1M & 2.9k & 2.9k
\end{tblr}
\end{table}

\begin{table}
\centering
\footnotesize
\caption{CNNs' throughput when deployed to GAP8 at its maximum performance configuration. The energy per inference is reported for two configurations ($E_{ee}$, $E_{mp}$). \blue{The last row corresponds to Tiny-PULP-Dronet v3.}}
\label{tab:network_onboard_performance}
\begin{tabular}{cccccccc} 
\toprule
\textbf{Blocks} & \textbf{by-pass} & $\bm{\gamma}$ & \textbf{Cycles} & $\bm{\frac{MAC}{Cycle}}$ & \textbf{frame/s} & \begin{tabular}[c]{@{}c@{}}$\bm{E_{ee}}$ \\\textbf{[mJ]}\end{tabular} & \begin{tabular}[c]{@{}c@{}}$\bm{E_{mp}}$ \\\textbf{[mJ]}\end{tabular} \\ 
\midrule
D+P & no & /1 & 5.1M & 2.4 & 34 & 2.1 & 3.0 \\
D+P & no & /2 & 2.9M & 1.8 & 61 & 1.1 & 1.7 \\
D+P & no & /4 & 1.7M & 1.4 & 101 & 0.6 & 1.0 \\
D+P & no & /8 & 1.3M & 0.9 & 139 & 0.4 & 0.7 \\
\bottomrule
\end{tabular}
\end{table}

\subsection{Quantization and deployment}\label{sec:quantization_deployment}

In this section, we progress with quantizing and deploying the four D+P models without by-passes, introduced in Section~\ref{subsec:size_choice}, to evaluate their trade-off between accuracy/RMSE and onboard execution efficiency.
We apply 8-bit post-training quantization, as detailed in Section~\ref{sec:quantization_deployment}.
Table~\ref{tab:quantized_net} outlines the quantized models' regression/classification, compared to non-quantized models in Table~\ref{tab:architecture_choice_size}.
Quantization introduces only $2\%$ reduction in accuracy on the CNN employing $\gamma=2$, while other CNNs keep the same score of the 32-bit precision.
Instead, the RSME is more sensitive to the new int8 data type, with the error slightly increasing of 0.011, 0.006, 0.005, and 0.009, for $\gamma=1,2,4,8$, respectively.
Compared to the original float32 models (Table~\ref{tab:architecture_choice_size}), we reduce by 4$\times$ the memory footprint of each CNN.

Then, we deploy these four quantized CNNs on the GAP8 SoC to analyze their on-device performances. 
Table~\ref{tab:network_onboard_performance} shows their inference rate (frame/s) when GAP8 is running at its \textit{max performance (mp)} configuration, i.e., FC@\SI{250}{\mega\hertz}, CL@\SI{175}{\mega\hertz}, and $V_{dd}$@\SI{1.2}{\volt} .
Our largest CNN model ($\gamma=1$) achieves a throughput of \SI{34}{\fps}, which is $1.8\times$ higher than the SoA PULP-Dronet v2, peaking at \SI{19}{\fps}, despite having the same number of channels across the CNN architecture; this improvement derives from our architecture modifications.
Other configurations of $\gamma$ result in \SI{61}{\fps} with \SI{14}{\kilo\byte}, \SI{101}{\fps} with \SI{4.7}{\kilo\byte}, and \SI{139}{\fps} with \SI{1.9}{\kilo\byte} for $\gamma = [2, 4, 8]$, respectively.
Our smallest model ($\gamma = 8$), called Tiny-PULP-Dronet v3, improves the throughput by  $7.3\times$ and reduces the memory footprint by $168\times$ compared to PULP-Dronet v2.

Last, we measure the power consumption of all CNNs under SoC configurations: \textit{i}) the max performance setting, and the energy-efficient (ee) one, which operates at FC@\SI{50}{\mega \hertz}, CL@\SI{100}{\mega \hertz}, and $V_{dd}$@\SI{1.0}{\volt}.
In the \textit{ee} configuration, the CNN with $\gamma \in {1,2}$ consume \SI{38}{\milli\watt}, while the CNNs with $\gamma \in {4,8}$ show an average power consumption of \SI{34}{\milli\watt}.
On the other hand, the \textit{mp} setting sees all CNNs averaging a power consumption of \SI{100}{\milli\watt}.
As shown in Table~\ref{tab:network_onboard_performance}, the energy for one-frame inference with the \textit{ee} configuration ($E_{ee}$) is \SI{2.1}{\nothing}, \SI{1.1}{\nothing}, \SI{0.6}{\nothing}, \SI{0.4}{\milli\joule} for the $\gamma = {1,2,4,8}$, respectively.
On the other hand, the one-frame inference energy needed for the \textit{mp} configuration ($E_{mp}$) is always $\sim1.5\times$ higher.

\begin{figure*}[tb]
\centering
\includegraphics[width=1.0\linewidth]{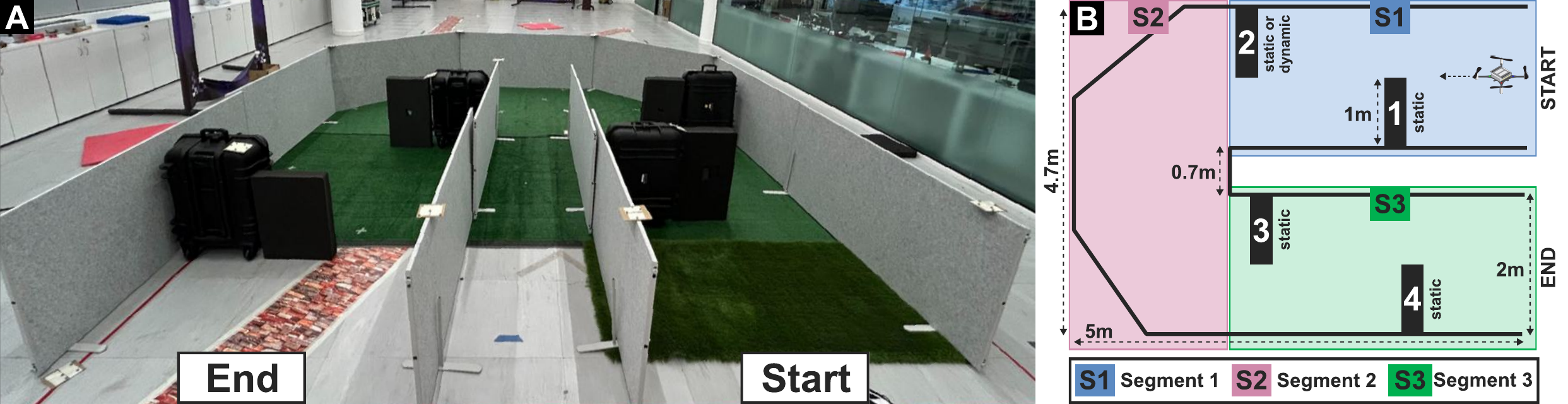}
\caption{A) our U-shaped path for the in-field experiments. B) T op-view 2D representation of the path, highlighting its division into three segments (S1, S2, and S3) and the position of the four obstacles (represented with black rectangles).}
\label{fig:u_shaped_path}
\end{figure*}

\section{In-field testing} \label{sec:infield_testing}

We evaluate the navigation capabilities of our PULP-Dronet v3 and Tiny-PULP-Dronet v3 CNNs with in-field experiments.
We record all experiments in a flying room equipped with a Qualisys motion capture system (24 cameras) with mm-precise tracking of our drone.
We track the position and pose of our drone @\SI{100}{\hertz} to analyze the drone’s flight in post-processing.
We investigate if the new dataset we collected, having unified labels for \textit{yaw-rate} and \textit{collision probability}, improves the navigation capabilities of the SoA PULP-Dronet v2, which was trained with disjoint \textit{steering} and \textit{collision probability} labels, and therefore struggles in avoiding static obstacles, as described in~\cite{pulpdronetv2JETCAS}.

We devised a challenging navigation scenario involving a U-shaped corridor, illustrated in Figure~\ref{fig:u_shaped_path}, and divided it into three segments. 
Segments S1 and S3 are straight paths, each one presenting two obstacles, whereas segment S2 features a 180-degree turn.
The \SI{2}{\meter} wide corridor has obstacles \SI{1}{\meter} wide that are leaning on opposite walls, blocking straight pathways.
We designed two experiments: \textit{i}) one where all obstacles are static, and \textit{ii}) one where the second one is dynamic.
In the experiment with a dynamic obstacle, obstacle 2 appears in the center of the lane as soon as the drone passes the first obstacle, providing \SI{1.5}{\meter} of braking space, and it is removed \SI{5}{\second} later.

\begin{figure*}
    \centering
    \includegraphics[width=1\linewidth]{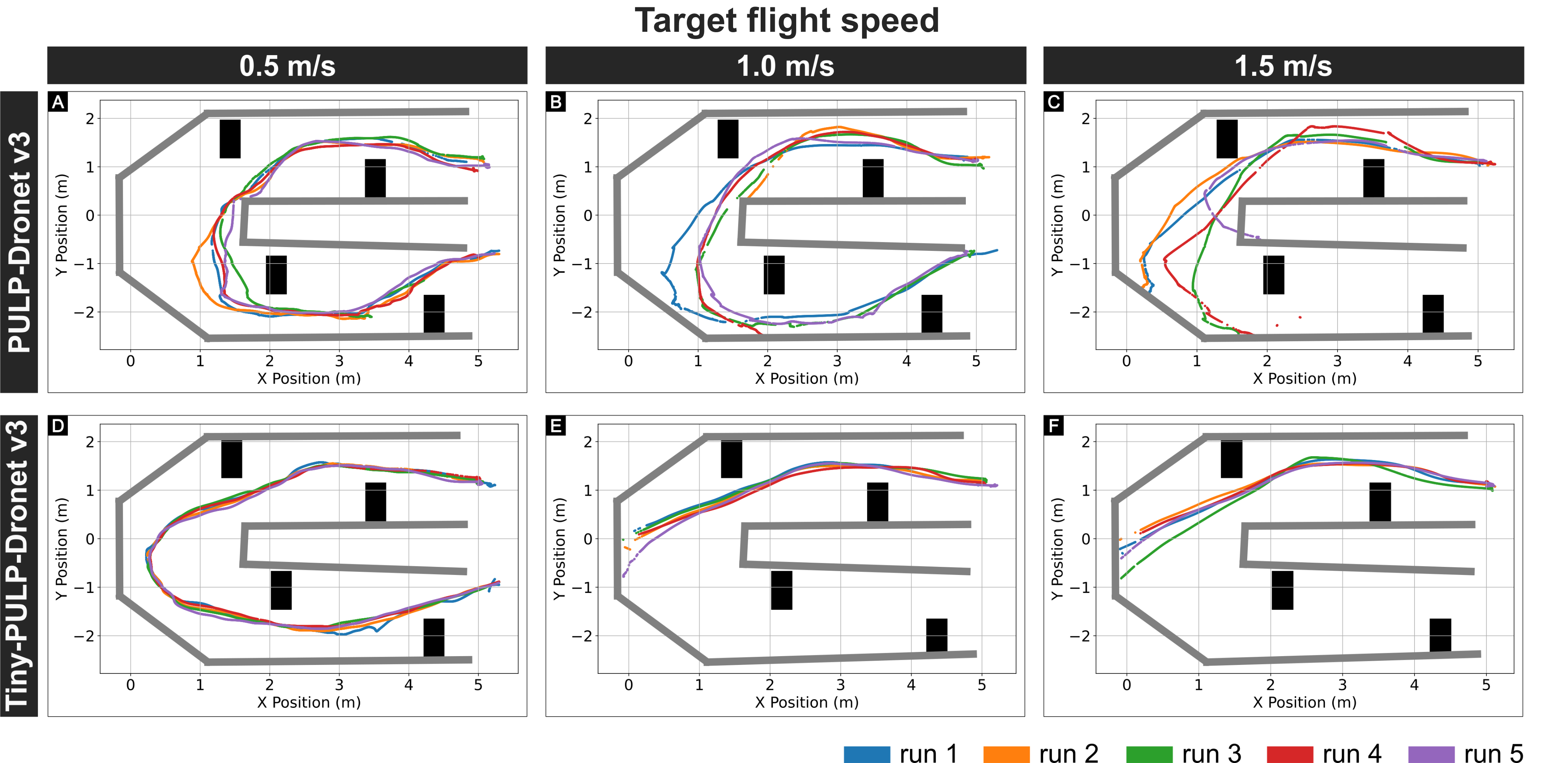}
    \caption{In-field experiments trajectories of PULP-Dronet v3 (A-B-C) and Tiny-PULP-Dronet v3 (D-E-F) for three target speeds $v_{target}=[0.5,1.0,1.5] $\SI{}{\meter/\second}, tested with static obstacles only (represented as black rectangles).}
    \label{fig:trajectories_static}
\end{figure*}

\begin{table*}
    \centering
    \caption{Success rate of our closed-loop system in the U-shaped path with static obstacles only. We report the success rate for the S1-S2-S3 segments, and the $v_{avg}$ over the complete path for PULP-Dronet v2, PULP-Dronet v3, and Tiny-PULP-Dronet v3.}
    \label{tab:results_static_obstacles}
    \begin{tblr}{
      column{3} = {c},
      column{4} = {c},
      column{5} = {c},
      column{6} = {c},
      column{8} = {c},
      column{9} = {c},
      column{10} = {c},
      column{13} = {c},
      column{14} = {c},
      column{15} = {c},
      cell{1}{1} = {c},
      cell{1}{3} = {c=4}{},
      cell{1}{8} = {c=4}{},
      cell{1}{13} = {c=4}{},
      cell{2}{3} = {c=3}{},
      cell{2}{8} = {c=3}{},
      cell{2}{13} = {c=3}{},
      cell{3}{11} = {c},
      cell{3}{16} = {c},
      cell{4}{1} = {c},
      cell{4}{11} = {c},
      cell{4}{16} = {c},
      cell{5}{1} = {c},
      cell{5}{11} = {c},
      cell{5}{16} = {c},
      cell{6}{1} = {c},
      cell{6}{11} = {c},
      cell{6}{16} = {c},
      hline{1,7} = {1,3-6,8-11,13-16}{0.08em},
      hline{2} = {1,3-6,8-11,13-16}{0.03em},
      hline{4} = {3-6,8-11,13-16}{0.03em},
    }
    \textbf{$\mathbf{v_{target} [m/s]}$} &  & \textbf{PULP-Dronet v2~\cite{pulpdronetv2JETCAS}} &  &  &  &  & \textbf{PULP-Dronet v3 (ours)} &  &  &  &  & \textbf{Tiny-PULP-Dronet v3 (ours)} &  &  & \\
     &  & success rate &  &  &  &  & success rate &  &  &  &  & success rate &  &  & \\
     &  & (S1) & (S2) & (S3) & $v_{avg}$ [\SI{}{\meter/\second}] &  & (S1) & (S2) & (S3) & $v_{avg}$ [\SI{}{\meter/\second}] &  & (S1) & (S2) & (S3) & $v_{avg}$ [\SI{}{\meter/\second}] \\
    0.5 &  & 0/5 & --- & --- & N/A &  & 5/5 & 5/5 & 4/5 & 0.39 &  & 5/5 & 5/5 & 5/5 & 0.45\\
    1   &  & 0/5 & --- & --- & N/A &  & 5/5 & 4/5 & 3/5 & 0.24 &  & 5/5 & 0/5 & --- & N/A \\
    1.5 &  & 0/5 & --- & --- & N/A &  & 5/5 & 0/5 & --- & N/A &  & 5/5 & 0/5 & --- & N/A
    \end{tblr}
\end{table*}

These scenarios aim at testing the nano-UAV's capabilities on multiple skills: \textit{i}) avoiding both static and dynamic obstacles, \textit{ii}) navigation through a narrow environment (a corridor), and \textit{iii}) a sharp \SI{180}{\degree} turn.
The corridor environment we use in our tests is \textit{never-seen-before} for both PULP-Dronet v2 and PULP-Dronet v3, not part of either CNN's training set.

We must stress that we designed these tests to challenge PULP-Dronet v2, focusing on its weaknesses specifically. 
While the CNN has demonstrated excellent navigation capabilities in corridor turns and dynamic obstacle avoidance, the original tests revealed that it struggles with static obstacles in narrow tunnels~\cite{pulpdronetv2JETCAS}.

We evaluate our closed-loop system across three specific target speeds: $v_{target}$ = \SI{0.5}{\nothing}, \SI{1.0}{\nothing}, \SI{1.5}{\nothing} \SI{}{\meter/\second}.
We set a target flight altitude of \SI{0.5}{\meter} in every test.
We conduct five experiments for each combination of CNN and $v_{target}$ for statistical relevance (total 90 tests), and we compare them with the SoA PULP-Dronet v2~\cite{pulpdronetv2JETCAS}.
For the sake of comparability, we always use the same drone's control state machine of PULP-Dronet v2~\cite{pulpdronetv2JETCAS}, including a first-order low-pass filter on both the forward speed of the drone $v_{unfilt}$ (Eq. \ref{eq:low_pass_filter_vel}) and yaw rate $\omega_{unfilt}$ (Eq. \ref{eq:low_pass_filter_yawrate}).
The filtered values for the forward velocity ($v_{filter}$) and the yaw rate  ($\omega_{filter}$) are fed to the drone's flight controller.
We set the low-pass filtering parameters $\alpha_1 = \alpha_2 = 0.3$.
We use the stock hardware from Bitcraze, including \textit{i}) stock motors (max current of \SI{1}{\ampere}), \textit{ii}) a \SI{250}{\milli\ampere\hour} battery, and \textit{iii}) stock propellers with a \SI{45}{\milli\meter} diameter, for a total weight of \SI{35}{\gram}.

The videos of the experiments are accessible through the link provided in the supplementary material section.

\begin{align}    % Use "&" at the beginning of the eq to left align.
    &v\sb{filter}(k) \leftarrow \alpha\sb{1} \cdot v\sb{unfilt} + (1 - \alpha\sb{1}) \cdot v\sb{filter}(k-1)
    \label{eq:low_pass_filter_vel}  % Equation 1
    \\ 
    &\omega\sb{filter}(k) \leftarrow \alpha\sb{2} \cdot \omega\sb{unfilt} + (1 - \alpha\sb{2}) \cdot \omega\sb{filter}(k-1)
    \label{eq:low_pass_filter_yawrate} % Equation 2
\end{align}

\subsection{U-shaped path with static obstacles} \label{subsec:infield_static_obstacles}
We conduct the first set of experiments with all static obstacles.
Table~\ref{tab:results_static_obstacles} outlines the success rate for each CNN tested along with the average speed ($v_{avg}$) of the drone.
If the drone fails to complete the entire path, $v_{avg}$ is noted as N/A. 
If it fails to complete a segment, we do not start the drone again on subsequent segments, and their success rates are noted as "---" when they are never attempted.
PULP-Dronet v2 never succeeds at any speed point to pass segment S1 of the path. 
The two obstacles of segment S1 visually obstruct the entire corridor's width, resulting in a consistently high collision probability, while the CNN's steering output keeps the drone in the center of the lane defined by the surrounding walls.
As a result, the drone either does not move forward due to the CNN's high collision probability, or it slowly drifts against obstacle 1, ultimately crashing.
This outcome aligns with a similar scenario described in~\cite{pulpdronetv2JETCAS}, where PULP-Dronet v2 encountered a $100\%$ failure rate in tackling a narrow tunnel scenario with static obstacles.

Moving to our CNNs, we plot the trajectories of PULP-Dronet v3 in Figure~\ref{fig:trajectories_static}-A-B-C, and the trajectories of Tiny-PULP-Dronet v3 in Figure~\ref{fig:trajectories_static}-D-E-F.
First, we analyze the results with a target speed of \SI{0.5}{\meter/\second}.
PULP-Dronet v3 and Tiny-PULP-Dronet v3 fly through the whole U-shaped path with a single failure and no failures, respectively. 
The only time that PULP-Dronet v3 fails, it crashes against obstacle 4, still successfully completing $66\%$ of the corridor.

Analyzing the @\SI{1}{\meter/\second} target speed configuration, PULP-Dronet v3 successfully navigates the entire corridor three times while crashing one time in segment S2 and once in segment S3.
On the other hand, Tiny-PULP-Dronet v3 reliably succeeds 5/5 times only in segment S1 of the path, always crashing during the turn of segment S2.
This outcome is attributed to Tiny-PULP-Dronet v3's lower RMSE performance compared to PULP-Dronet v3, as detailed in Section~\ref{sec:quantization_deployment}.
Analyzing the @\SI{1.5}{\meter/\second} target speed configuration, none of the CNNs tested succeeded in completing the turn, outlining the speed upper limit of our closed-loop systems in this scenario, but still succeeding $100\%$ of the times in avoiding the obstacles in segment S1.

In conclusion, the static obstacle avoidance success rate of PULP-Dronet v3 and Tiny-PULP-Dronet v3 is inversely proportional to the target flight speed. 
While PULP-Dronet v3 and Tiny-PULP-Dronet v3 show $80\%$ and $100\%$ success rates at \SI{0.5}{\meter/\second}, their success rates over the whole path lowers at higher speeds.
PULP-Dronet v3 reduces its success rate to $60\%$ at \SI{1}{\meter\second}, while Tiny-PULP-Dronet v3 and PULP-Dronet v3 never succeed at \SI{1}{\meter\second} and \SI{1.5}{\meter\second}, respectively.
These results mark an improvement with respect to PULP-Dronet v2, showing that \textit{i}) the dataset we collected with joint labels successfully trains CNNs that can tackle both obstacle avoidance and steering tasks together,  and \textit{ii}) our Tiny-PULP-Dronet v3, with only \SI{2.9}{\kilo\nothing} parameters, can enable static obstacle avoidance while being $168\times$ smaller than the SoA model.

\begin{figure*}[t]
    \centering
    \includegraphics[width=1\linewidth]{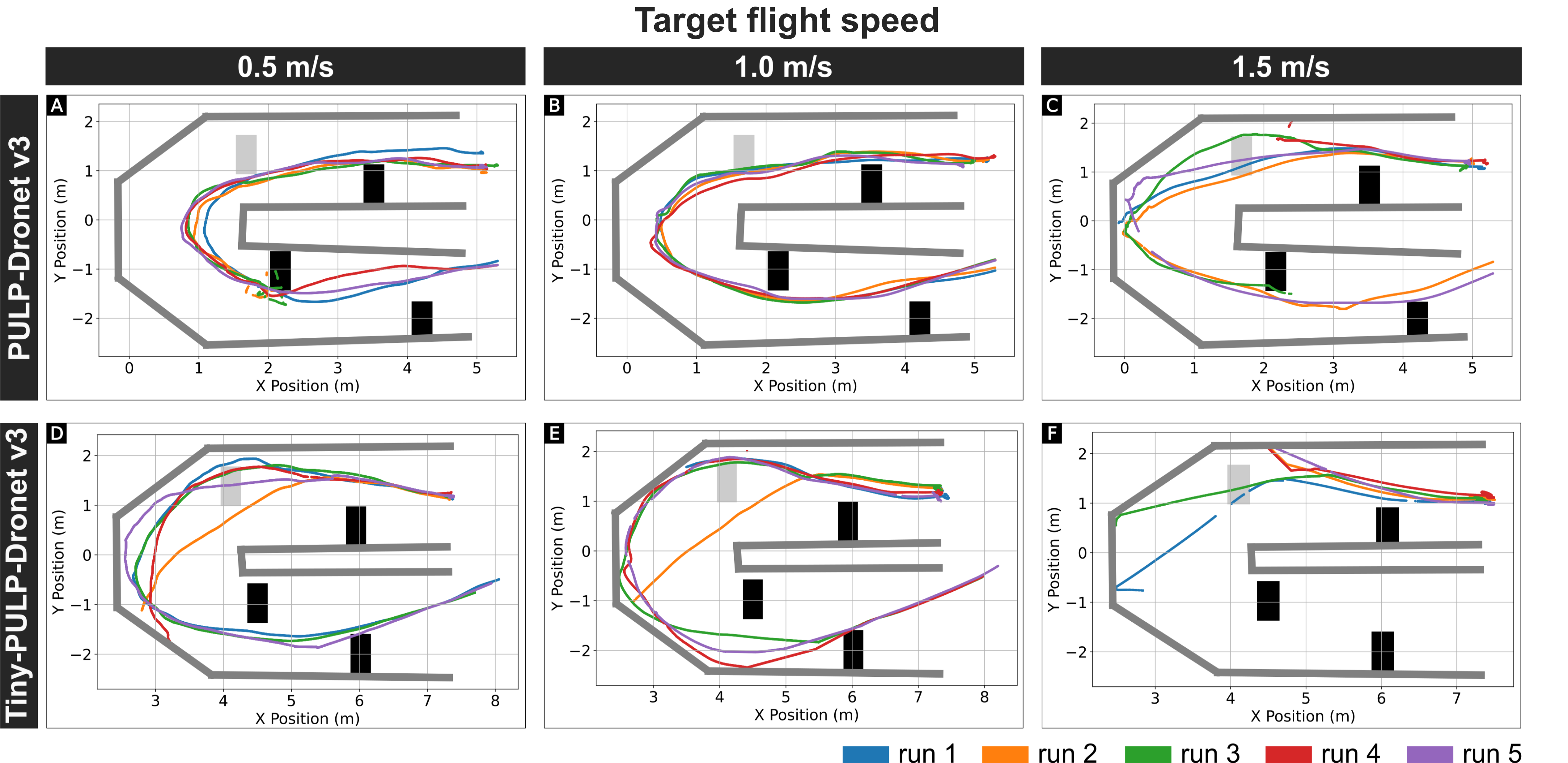}
    \caption{In-field experiments trajectories of PULP-Dronet v3 (A-B-C) and Tiny-PULP-Dronet v3 (D-E-F) for three target speeds $v_{target}=0.5, 1.0, 1.5 $\SI{}{\meter/\second}, tested with three static obstacles (represented as black rectangles) and a dynamic one (represented as a light grey rectangle). }
    \label{fig:trajectories_static_dynamic_obstacles}
\end{figure*}

\begin{table*}
\centering
\caption{Success rate of our closed-loop system in the U-shaped path with three static obstacles and a dynamic one. We report the success rate for the S1-S2-S3 segments, and the $v_{avg}$ over the complete path for PULP-Dronet v2, PULP-Dronet v3, and Tiny-PULP-Dronet v3.}
\label{tab:results_static_and_dynamic_obstacles}
\begin{tblr}{
  column{3} = {c},
  column{4} = {c},
  column{5} = {c},
  column{6} = {c},
  column{8} = {c},
  column{9} = {c},
  column{10} = {c},
  column{13} = {c},
  column{14} = {c},
  column{15} = {c},
  cell{1}{1} = {c},
  cell{1}{3} = {c=4}{},
  cell{1}{8} = {c=4}{},
  cell{1}{13} = {c=4}{},
  cell{2}{3} = {c=3}{},
  cell{2}{8} = {c=3}{},
  cell{2}{13} = {c=3}{},
  cell{3}{11} = {c},
  cell{3}{16} = {c},
  cell{4}{1} = {c},
  cell{4}{11} = {c},
  cell{4}{16} = {c},
  cell{5}{1} = {c},
  cell{5}{11} = {c},
  cell{5}{16} = {c},
  cell{6}{1} = {c},
  cell{6}{11} = {c},
  cell{6}{16} = {c},
  hline{1,7} = {1,3-6,8-11,13-16}{0.08em},
  hline{2} = {1,3-6,8-11,13-16}{0.03em},
  hline{4} = {3-6,8-11,13-16}{0.03em},
}
\textbf{$\mathbf{v_{target} [m/s]}$} &  & \textbf{PULP-Dronet v2~\cite{pulpdronetv2JETCAS}} &  &  &  &  & \textbf{PULP-Dronet v3 (ours)} &  &  &  &  & \textbf{Tiny-PULP-Dronet v3 (ours)} &  &  & \\
 &  & success rate &  &  &  &  & success rate &  &  &  &  & success rate &  &  & \\
 &  & (S1) & (S2) & (S3) & $v_{avg}$ [\SI{}{\meter/\second}] &  & (S1) & (S2) & (S3) & $v_{avg}$ [\SI{}{\meter/\second}] &  & (S1) & (S2) & (S3) & $v_{avg}$ [\SI{}{\meter/\second}]\\
0.5 &  & 0/5 & --- & --- & N/A &  & 5/5 & 5/5 & 3/5 & 0.57 &  & 5/5 & 5/5 & 3/5 & 0.44 \\
1   &  & 0/5 & --- & --- & N/A &  & 5/5 & 5/5 & 5/5 & 0.98 &  & 5/5 & 5/5 & 3/5 & 0.82\\
1.5 &  & 0/5 & --- & --- & N/A &  & 4/5 & 3/5 & 2/5 & 1.33 &  & 2/5 & 0/5 & --- & N/A
\end{tblr}
\end{table*}

\subsection{U-shaped path with static and dynamic obstacles} \label{subsec:infield_static_dynamic_obstacles}

We conduct the second set of experiments stressing dynamic obstacle avoidance. 
We use the same setup as Section~\ref{subsec:infield_static_obstacles}, but now obstacle 2 appears in the center of the corridor after the drone passes obstacle 1. 
Table~\ref{tab:results_static_and_dynamic_obstacles} outlines the success rate of the three CNNs tested.
Starting from the SoA CNN, PULP-Dronet v2 never succeeds in navigating any of the three segments (S1, S2, S3) of our path for the same reason described in Section~\ref{subsec:infield_static_obstacles}, getting stuck in front of obstacle 1 and eventually crashing into it.

Moving to our CNNs, we plot the trajectories of PULP-Dronet v3 in Figure~\ref{fig:trajectories_static_dynamic_obstacles}-A-B-C, and the trajectories of Tiny-PULP-Dronet v3 in Figure~\ref{fig:trajectories_static_dynamic_obstacles}-D-E-F. 
PULP-Dronet v3 shows the highest success rate among all the CNNs tested.
It completes the whole path with a success rate of $60\%$, $100\%$, $40\%$ when flying at 0.5, 1.0, 1.5\SI{}{\meter/\second}, respectively.
Remarkably, PULP-Dronet v3 only crashed once against the dynamic obstacle while flying at the highest speed (\SI{1.5}{\meter/\second}), always succeeding in passing the S1 segment in all other cases.
On the other hand, the Tiny-PULP-Dronet v3 completes the whole corridor with a $60\%$ success rate when flying at both \SI{0.5}{\meter/\second} and \SI{1}{\meter/\second}.
However, it always fails at a higher speed of \SI{1.5}{\meter/\second}, either crashing on the right wall of segment S1 or failing to complete the turn in S2.
Nevertheless, Tiny-PULP-Dronet v3 avoided the collision against the dynamic obstacle of segment S1 two times when flying at the highest speed.

In conclusion, our experiments demonstrate that for PULP-Dronet v3 and Tiny-PULP-Dronet v3, the dynamic obstacle avoidance rate is inversely proportional to the drone’s flight speed when it exceeds \SI{1}{\meter/\second}. 
Both CNNs reliably avoid collisions with the dynamic obstacle in Section S1 $100\%$ of the time at speeds up to \SI{1}{\meter/\second}. 
However, at the highest speed of \SI{1.5}{\meter/\second}, the success rates decrease to $80\%$ for PULP-Dronet v3 and $40\%$ for Tiny-PULP-Dronet v3, respectively.

\section{Conclusion} \label{sec:conclusion}

Nano-sized UAVs are ideal candidates as ubiquitous flying IoT nodes.
In this paper, we distill a novel family of CNNs for autonomous navigation on nano-drones, i.e., the Tiny-PULP-Dronet v3 CNNs.
Compared to the SoA, our models reduce memory footprint by up to 168$\times$ (down to \SI{2.9}{\kilo\byte}) and achieve an inference rate of up to \SI{139}{\fps}. 
We create a new open-source \SI{66}{\kilo\nothing} image dataset for autonomous nano-UAV navigation. 
We compare with in-field tests both SoA and our CNNs on a COTS nano-UAV. 
Our tiny CNN succeeds in navigating a challenging path with static and dynamic obstacles and a \SI{180}{\degree} turn at speeds of up to \SI{1}{\meter/\second}. 
In contrast, the baseline consistently fails despite having 168$\times$ more parameters.

For future development, we aim to exploit the compute and memory budget we have freed with Tiny-PULP-Dronet v3 by introducing additional tasks,  e.g., executing multiple CNNs onboard~\cite{tiny_dronets}. 
Additionally, efficiency can be further improved by implementing a more aggressive quantization scheme~\cite{quantization_survey}, e.g., 4-bit or 2-bit, coupled with quantization-aware training, to speed up computation and reduce the memory footprint while maintaining accuracy.

\section*{Acknowledgment}
We thank the Autonomous Robotics Research Center of the Technology Innovation Institute, Abu Dhabi, UAE, for granting us access to their indoor flying arena.
We thank Micha\l{} Barci\'s and Daniel Ribien for their support in dataset collection, and Elia Cereda for his support and assistance.
% We thank Micha\l{} Barci\'s for helping in the dataset collection and Daniel Ribien for helping build the dataset collector framework.
% We thank Elia Cereda for his assistance.

\bibliographystyle{IEEEtran}
\bibliography{IEEEabrv,main}

% Generated by IEEEtran.bst, version: 1.14 (2015/08/26)
\begin{thebibliography}{10}
\providecommand{\url}[1]{#1}
\csname url@samestyle\endcsname
\providecommand{\newblock}{\relax}
\providecommand{\bibinfo}[2]{#2}
\providecommand{\BIBentrySTDinterwordspacing}{\spaceskip=0pt\relax}
\providecommand{\BIBentryALTinterwordstretchfactor}{4}
\providecommand{\BIBentryALTinterwordspacing}{\spaceskip=\fontdimen2\font plus
\BIBentryALTinterwordstretchfactor\fontdimen3\font minus \fontdimen4\font\relax}
\providecommand{\BIBforeignlanguage}[2]{{%
\expandafter\ifx\csname l@#1\endcsname\relax
\typeout{** WARNING: IEEEtran.bst: No hyphenation pattern has been}%
\typeout{** loaded for the language `#1'. Using the pattern for}%
\typeout{** the default language instead.}%
\else
\language=\csname l@#1\endcsname
\fi
#2}}
\providecommand{\BIBdecl}{\relax}
\BIBdecl

\bibitem{inclined_landing}
J.~E. Kooi and R.~Babu{\v s}ka, ``{{Inclined Quadrotor Landing Using Deep Reinforcement Learning}},'' in \emph{2021 {{IEEE}}/{{RSJ}} International Conference on Intelligent Robots and Systems ({{IROS}})}, 2021, pp. 2361--2368.

\bibitem{neural_swarm_2}
G.~Shi, W.~H{\"o}nig, X.~Shi, Y.~Yue, and S.-J. Chung, ``{{Neural-Swarm2: Planning and Control of Heterogeneous Multirotor Swarms Using Learned Interactions}},'' \emph{IEEE Transactions on Robotics}, pp. 1--17, 2021.

\bibitem{cereda_nas}
E.~Cereda, L.~Crupi, M.~Risso, A.~Burrello, L.~Benini, A.~Giusti, D.~Jahier~Pagliari, and D.~Palossi, ``{{Deep Neural Network Architecture Search for Accurate Visual Pose Estimation Aboard Nano-UAVs}},'' in \emph{{{2023 {{IEEE}} International Conference on Robotics and Automation (ICRA)}}}, 2023, pp. 6065--6071.

\bibitem{lamberti_exploration_detection}
L.~Lamberti, L.~Bompani, V.~J. Kartsch, M.~Rusci, D.~Palossi, and L.~Benini, ``{{Bio-inspired Autonomous Exploration Policies with CNN-based Object Detection on Nano-drones}},'' in \emph{2023 Design, Automation \& Test in Europe Conference \& Exhibition ({{DATE}})}, 2023, pp. 1--6.

\bibitem{pulpdronetv2JETCAS}
V.~Niculescu, L.~Lamberti, F.~Conti, L.~Benini, and D.~Palossi, ``{{Improving Autonomous Nano-Drones Performance via Automated End-to-End Optimization and Deployment of DNNs}},'' \emph{IEEE Journal on Emerging and Selected Topics in Circuits and Systems}, pp. 1--1, 2021.

\bibitem{uavs_iot_survey}
N.~S. Labib, M.~R. Brust, G.~Danoy, and P.~Bouvry, ``{{The Rise of Drones in Internet of Things: A Survey on the Evolution, Prospects and Challenges of Unmanned Aerial Vehicles}},'' \emph{{{IEEE Access}}}, vol.~9, pp. 115\,466--115\,487, 2021.

\bibitem{uav_iot_sensing}
Z.~Wei, M.~Zhu, N.~Zhang, L.~Wang, Y.~Zou, Z.~Meng, H.~Wu, and Z.~Feng, ``{{UAV-Assisted Data Collection for Internet of Things: A Survey}},'' \emph{IEEE Internet of Things Journal}, vol.~9, no.~17, pp. 15\,460--15\,483, 2022.

\bibitem{higgins_pathplanning_cluttered}
J.~Higgins, N.~Mohammad, and N.~Bezzo, ``{{A Model Predictive Path Integral Method for Fast, Proactive, and Uncertainty-Aware UAV Planning in Cluttered Environments}},'' in \emph{{{2023 IEEE/RSJ International Conference on Intelligent Robots and Systems (IROS)}}}, 2023, pp. 830--837.

\bibitem{cereda2021improving}
E.~Cereda, M.~Ferri, D.~Mantegazza, N.~Zimmerman, L.~M. Gambardella, J.~Guzzi, A.~Giusti, and D.~Palossi, ``{{Improving the Generalization Capability of {{DNNs}} for Ultra-Low Power Autonomous Nano-UAVs}},'' in \emph{2021 17th International Conference on Distributed Computing in Sensor Systems ({{DCOSS}})}, 2021, pp. 327--334.

\bibitem{uav_chemicals}
M.~J. Anderson, J.~G. Sullivan, J.~L. Talley, K.~M. Brink, S.~B. Fuller, and T.~L. Daniel, ``{{The ``Smellicopter,'' a Bio-Hybrid Odor Localizing Nano Air Vehicle}},'' in \emph{2019 {{IEEE}}/{{RSJ}} International Conference on Intelligent Robots and Systems ({{IROS}})}, 2019, pp. 6077--6082.

\bibitem{uav_iot_survey}
N.~Hossein~Motlagh, T.~Taleb, and O.~Arouk, ``{{Low-Altitude Unmanned Aerial Vehicles-Based Internet of Things Services: Comprehensive Survey and Future Perspectives}},'' \emph{IEEE Internet of Things Journal}, vol.~3, no.~6, pp. 899--922, 2016.

\bibitem{uav_survey_applications}
H.~Shakhatreh, A.~H. Sawalmeh, A.~Al-Fuqaha, Z.~Dou, E.~Almaita, I.~Khalil, N.~S. Othman, A.~Khreishah, and M.~Guizani, ``{{Unmanned Aerial Vehicles (UAVs): A Survey on Civil Applications and Key Research Challenges}},'' \emph{IEEE Access}, vol.~7, pp. 48\,572--48\,634, 2019.

\bibitem{zhang2024_depthestimation_slam}
N.~Zhang, F.~Nex, G.~Vosselman, and N.~Kerle, ``{{End-to-End Nano-Drone Obstacle Avoidance for Indoor Exploration}},'' \emph{{{Drones}}}, vol.~8, no.~2, p.~33, 2024.

\bibitem{drone_racing_survey}
D.~Hanover, A.~Loquercio, L.~Bauersfeld, A.~Romero, R.~Penicka, Y.~Song, G.~Cioffi, E.~Kaufmann, and D.~Scaramuzza, ``{{Autonomous Drone Racing: A Survey}},'' \emph{{{IEEE Transactions on Robotics}}}, vol.~40, pp. 3044--3067, 2024.

\bibitem{inverted_landing}
B.~Habas, J.~W. Langelaan, and B.~Cheng, ``{{Inverted Landing in a Small Aerial Robot via Deep Reinforcement Learning for Triggering and Control of Rotational Maneuvers}},'' in \emph{{{2023 IEEE International Conference on Robotics and Automation (ICRA)}}}, 2023, pp. 3368--3375.

\bibitem{garofalo2020pulpnn}
A.~Garofalo, M.~Rusci, F.~Conti, D.~Rossi, and L.~Benini, ``{{PULP-NN: Accelerating Quantized Neural Networks on Parallel Ultra-Low-Power RISC-V Processors}},'' \emph{Philosophical Transactions of the Royal Society A}, vol. 378, no. 2164, p. 20190155, 2020.

\bibitem{mobilenetv2}
M.~Sandler, A.~Howard, M.~Zhu, A.~Zhmoginov, and L.-C. Chen, ``{{MobileNetV2: Inverted Residuals and Linear Bottlenecks}},'' \emph{arXiv:1801.04381}, 2019.

\bibitem{tiny_dronets}
L.~Lamberti, V.~Niculescu, M.~Barci{\'s}, L.~Bellone, E.~Natalizio, L.~Benini, and D.~Palossi, ``{{Tiny-PULP-Dronets: Squeezing Neural Networks for Faster and Lighter Inference on Multi-Tasking Autonomous Nano-Drones}},'' in \emph{2022 {{IEEE}} 4th International Conference on Artificial Intelligence Circuits and Systems ({{AICAS}})}, 2022, pp. 287--290.

\bibitem{decroon_dataset}
J.~Dupeyroux, R.~Dinaux, N.~Wessendorp, and G.~De~Croon, ``{{A Novel Obstacle Detection and Avoidance Dataset for Drones}},'' in \emph{{{DroneSE and RAPIDO: System Engineering for constrained embedded systems}}}, 2022, pp. 8--13.

\bibitem{loquercio2018dronet}
A.~Loquercio, A.~I. Maqueda, C.~R. {Del-Blanco}, and D.~Scaramuzza, ``{{Dronet: Learning to Fly by Driving}},'' \emph{IEEE Robotics and Automation Letters}, vol.~3, no.~2, pp. 1088--1095, 2018.

\bibitem{drone_dataset_hdin}
Y.~Chang, Y.~Cheng, J.~Murray, S.~Huang, and G.~Shi, ``{{The {{HDIN}} Dataset: {{A}} Real-World Indoor {{UAV}} Dataset with Multi-Task Labels for Visual-Based Navigation}},'' \emph{Drones}, vol.~6, no. 202, 2022.

\bibitem{navardi_pulp_dronet}
M.~Navardi, E.~Humes, and T.~Mohsenin, ``{{E2EdgeAI: Energy-efficient Edge Computing for Deployment of Vision-Based DNNs on Autonomous Tiny Drones}},'' in \emph{2022 {{IEEE}}/{{ACM}} 7th Symposium on Edge Computing ({{SEC}})}, 2022, pp. 504--509.

\bibitem{quantization_survey}
T.~Liang, J.~Glossner, L.~Wang, S.~Shi, and X.~Zhang, ``{{Pruning and quantization for deep neural network acceleration: A survey}},'' \emph{Neurocomputing}, vol. 461, pp. 370--403, 2021.

\bibitem{conti2020nemo}
F.~Conti, ``{{Technical Report: {{NEMO DNN}} Quantization for Deployment Model}},'' \emph{arXiv:2004.05930}, 2020.

\bibitem{burrello2020dory}
A.~Burrello, A.~Garofalo, N.~Bruschi, G.~Tagliavini, D.~Rossi, and F.~Conti, ``{{DORY: Automatic End-to-End Deployment of Real-World DNNs on Low-Cost IoT MCUs}},'' \emph{{{IEEE Transactions on Computers}}}, vol.~70, no.~8, pp. 1253--1268, 2021.

\bibitem{resnet2022}
S.~S. Saini and P.~Rawat, ``Deep residual network for image recognition,'' in \emph{{{2022 IEEE International Conference on Distributed Computing and Electrical Circuits and Electronics (ICDCECE)}}}, 2022, pp. 1--4.

\bibitem{mobilenetv1}
A.~G. Howard, M.~Zhu, B.~Chen, D.~Kalenichenko, W.~Wang, T.~Weyand, M.~Andreetto, and H.~Adam, ``{{MobileNets: Efficient Convolutional Neural Networks for Mobile Vision Applications}},'' \emph{arXiv:1704.04861}, 2017.

\end{thebibliography}

\begin{IEEEbiography}[{\includegraphics[width=0.95in,height=1.25in,clip,keepaspectratio]{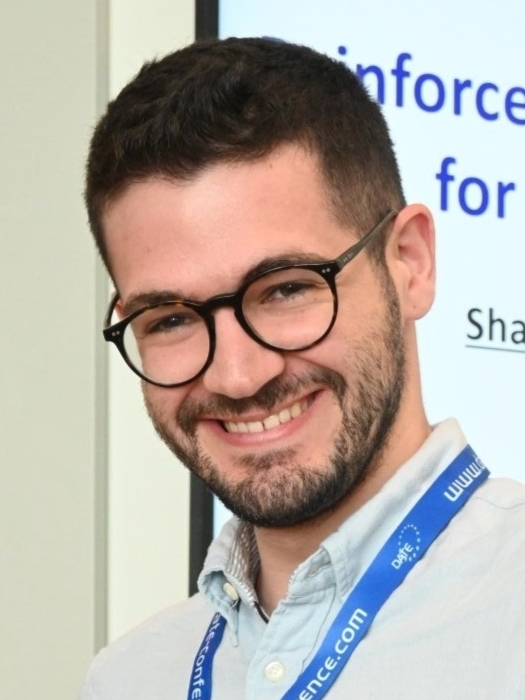}}]
{Lorenzo Lamberti} (Member, IEEE) received his Ph.D. in electronic engineering from the University of Bologna, Italy, in 2024, where he now holds a researcher position. 
His research encompasses artificial intelligence, miniaturized robotics, ultra-low-power embedded systems, and neuromorphic vision. 
His work has resulted in over 12 publications in international conferences and journals.
Dr. Lamberti was the technical lead of the winning team in the  ``Nanocopter AI Challenge,'' hosted at the IMAV'22 International Conference in TU Delft.
\end{IEEEbiography}

\begin{IEEEbiography}[{\includegraphics[width=0.95in,height=1.25in,clip,keepaspectratio]{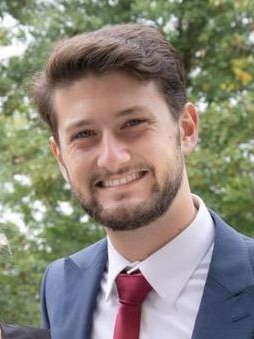}}]
{Lorenzo Bellone} received his B.Sc from the Faculty of Industrial Engineering and M.Sc from the Faculty of Telecommunication Engineering in 2018 and 2021, respectively, at Politecnico di Torino, Turin, Italy. 
He currently holds a Researcher position at the Autonomous Robotics Research Center with the Technology Innovation Institute, Abu Dhabi, UAE. 
His research interests focus on Deep Learning for communication networks, UAV-aided networks, and deployment of ML solutions for aerial multi-agent systems.
\end{IEEEbiography}

\begin{IEEEbiography}[{\includegraphics[width=0.95in,height=1.25in,clip,keepaspectratio]{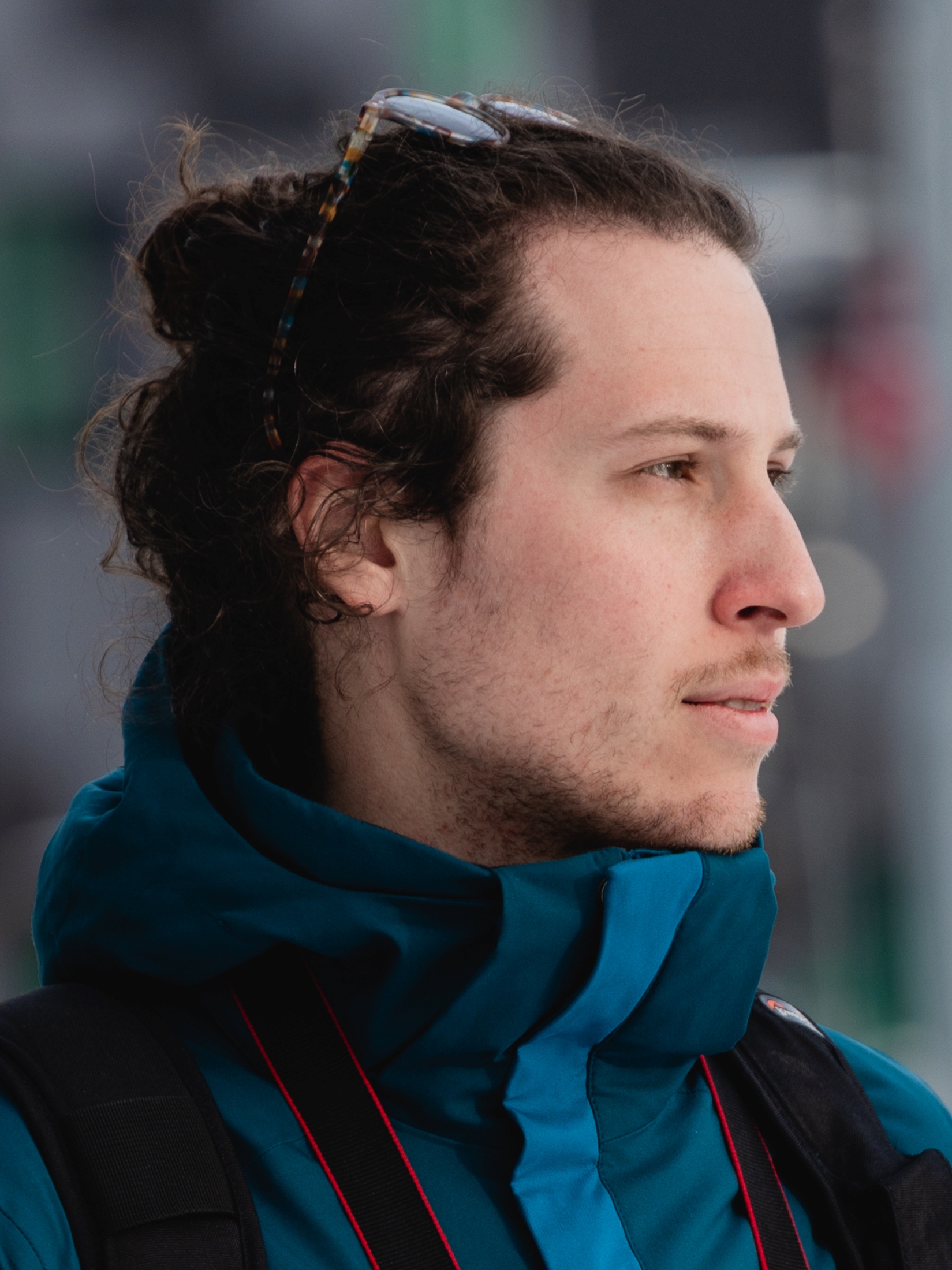}}]
{Luka Macan} received his B.Sc. and M.Sc. degrees from the Faculty of Electrical Engineering and Computing, University of Zagreb, Croatia, in 2017 and 2019, respectively. 
He is currently pursuing a Ph.D. degree at the University of Bologna. 
His research interests include machine learning on embedded systems and hardware accelerators.
\end{IEEEbiography}

\begin{IEEEbiography}[{\includegraphics[width=0.95in,height=1.25in,clip,keepaspectratio]{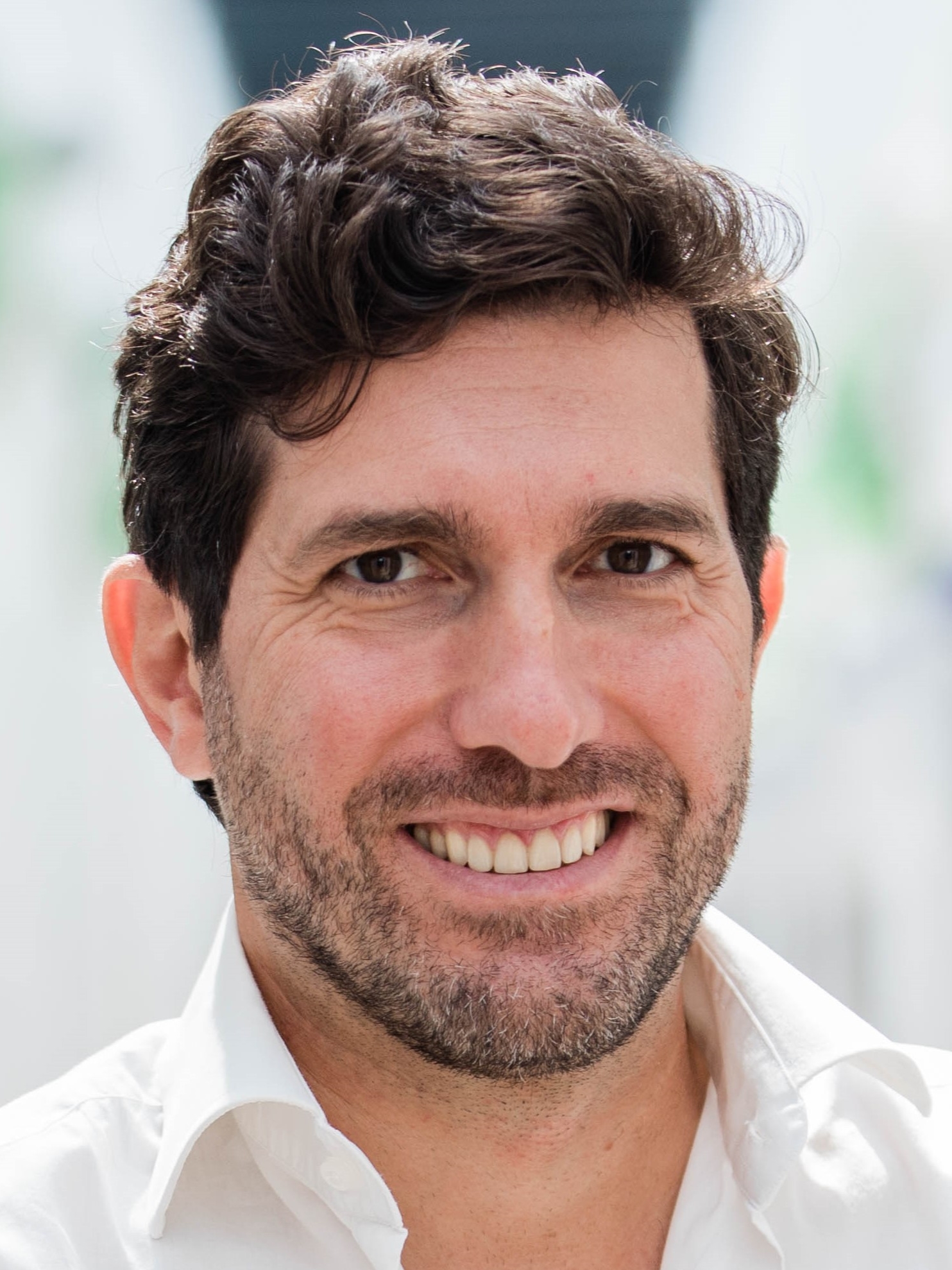}}]
{Enrico Natalizio} (Senior Member, IEEE) is currently the Chief Researcher of the Autonomous Robotics Research Center with the Technology Innovation Institute (UAE) and a Full Professor with the LORIA laboratory at the Université de Lorraine (France). 
He obtained his master’s degree magna cum laude and his Ph.D in Computer Engineering at the University of Calabria (Italy) in 2000 and 2005 respectively. 
In 2005-2006, he was a visiting researcher at the BWN (Broadband Wireless Networking) Lab at Georgia Tech in Atlanta (USA). 
From 2006 to 2010, he was a research fellow at the Titan Lab of the Università della Calabria (Italy). 
In October 2010, he joined POPS team at Inria Lille – Nord Europe (France) as a postdoc researcher and from 2012 till 2018 he was an Associate Professor at the Université de technologie de Compiègne (France), and Full Professor at the Université de Lorraine, from September 2018. 
His research interests include UAV communications and networking, robot and sensor communications with applications in networking technologies for disaster management and infrastructure monitoring, and IoT privacy and security. 
He is currently an associated editor of Elsevier Vehicular Communications, and Computer Networks. 
He has been ranked in the top 2\% worldwide scientists in Stanford University's bibliometric study of 2021.
\end{IEEEbiography}

\begin{IEEEbiography}[{\includegraphics[width=0.95in,height=1.25in,clip,keepaspectratio]{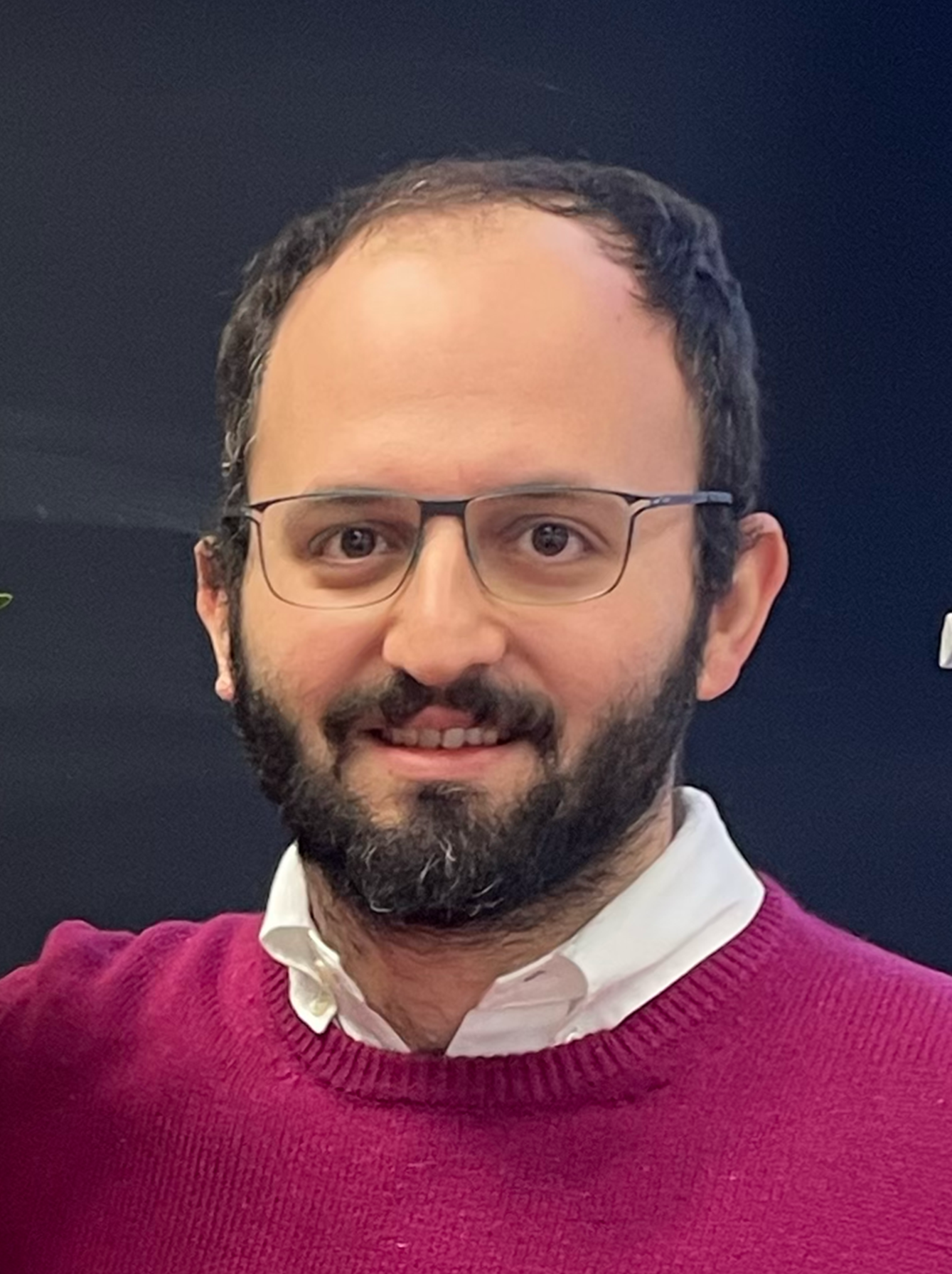}}]
{Francesco Conti} (Member, IEEE) received his Ph.D. degree in electronic engineering from the University of Bologna, Italy, in 2016.
He is currently a Tenure-Track Assistant Professor with the DEI Department at the University of Bologna. 
From 2016 to 2020, he held a research grant with the University of Bologna and a position as a Post-Doctoral Researcher with ETH Zürich.
His research is centered on hardware acceleration in ultra-low power and highly energy-efficient platforms, with a particular focus on System-on-Chips for Artificial Intelligence applications.
His research work has resulted in more than 90 publications in international conferences and journals and was awarded several times, including the 2020 IEEE \textsc{Transactions on Circuits and Systems I: Regular Papers} Darlington Best Paper Award.
\end{IEEEbiography}

\begin{IEEEbiography}[{\includegraphics[width=0.95in,height=1.25in,clip,keepaspectratio]{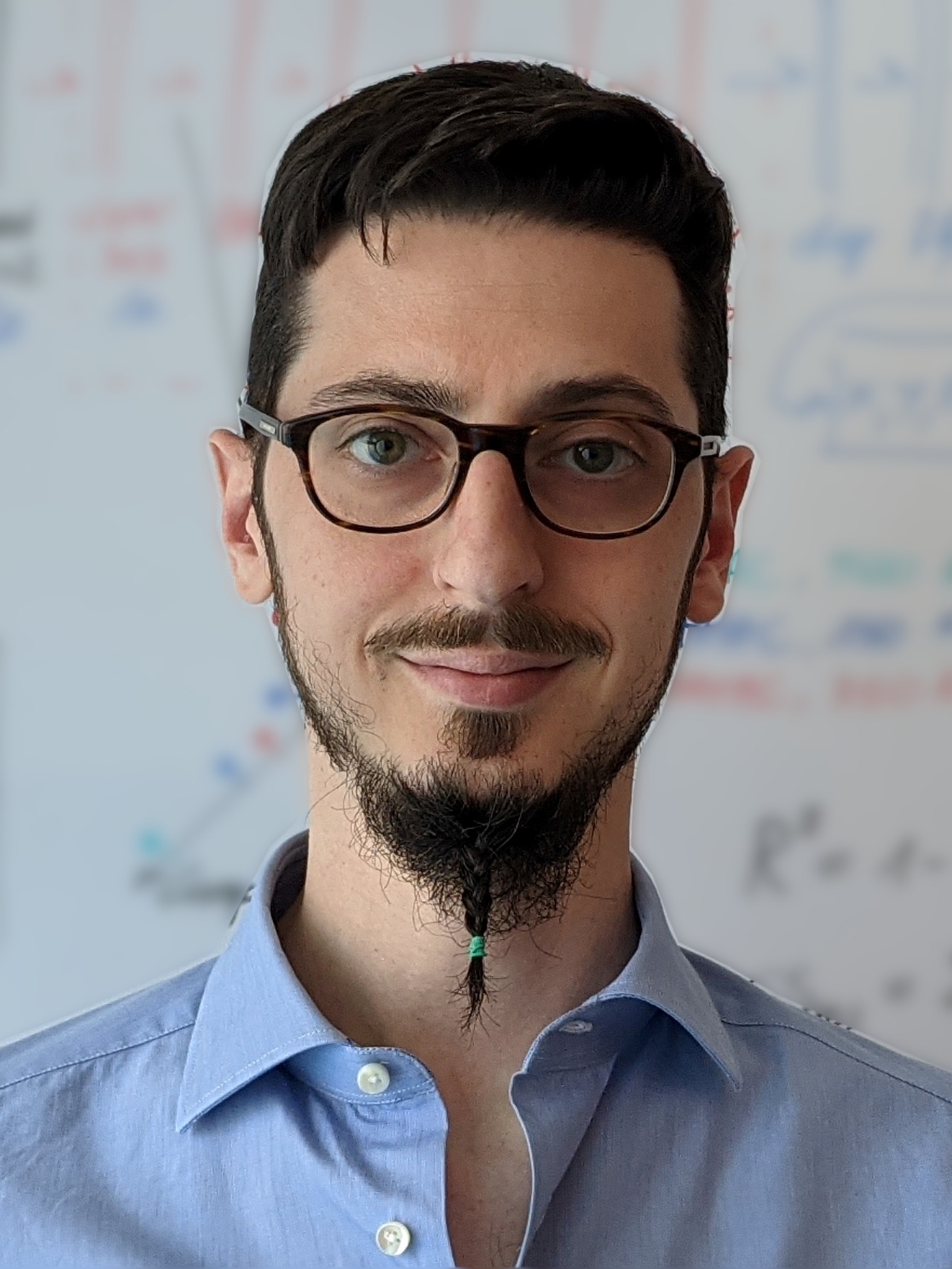}}]
{Daniele Palossi} (he/his) received his Ph.D. in Information Technology and Electrical Engineering from ETH Zürich. He is currently a Senior Researcher at the Dalle Molle Institute for Artificial Intelligence (IDSIA), USI-SUPSI, Lugano, Switzerland, where he leads the nano-robotics research group, and at the Integrated Systems Laboratory (IIS), ETH Zürich, Zürich, Switzerland. His research stands at the intersection of artificial intelligence, ultra-low-power embedded systems, and miniaturized robotics. His work has resulted in 45+ peer-reviewed publications in international conferences and journals. Dr. Palossi was a recipient of the Swiss National Science Foundation (SNSF) Spark Grant, the 2nd prize at the Design Contest held at the ACM/IEEE ISLPED'19, several Best Paper Awards, and team leader of the winning team of the first ``Nanocopter AI Challenge'' hosted at the IMAV'22 International Conference.
\end{IEEEbiography}

\begin{IEEEbiography}[{\includegraphics[width=0.95in,height=1.25in,clip,keepaspectratio]{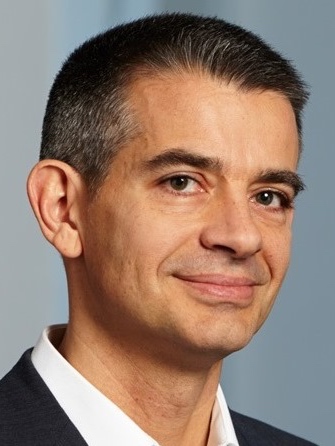}}]
{Luca Benini} (Fellow, IEEE) holds the chair of digital Circuits and systems at ETHZ and is a Full Professor at the Universit\`a di Bologna. 
He received a PhD from Stanford University. 
His research interests are in energy-efficient parallel computing systems and machine learning hardware. 
He is a Fellow of the ACM and a member of the Academia Europaea. 
He is the recipient of the 2016 IEEE CAS Mac Van Valkenburg Award, the 2020 EDAA Achievement Award, and the 2020 ACM/IEEE A. Richard Newton Award and the 2023 IEEE CS E.J. McCluskey Award.
\end{IEEEbiography}

\end{document}